\documentclass[runningheads,english,a4paper,10pt]{article}%
\pdfoutput=1
\usepackage[utf8]{inputenc}%
\usepackage{babel}%
\usepackage{amsmath,amssymb,amsfonts}
\usepackage{theorem}%

\theoremstyle{definition}%
\usepackage{bm}
\usepackage{mathrsfs}
\usepackage{eufrak}%
\usepackage[sort&compress]{natbib}
\usepackage{graphicx}%
\usepackage{txfonts}%
\usepackage{paralist}%
\usepackage{varioref}%
\usepackage{textcomp}%
\usepackage[T1]{fontenc}%
\usepackage{lmodern}%
\usepackage{wasysym}
\usepackage{pict2e}%
\newcommand{\sub}[2]{\ensuremath{{#1}_{_{#2}}}}%
\newcommand{\eN}{\ensuremath{\sub{\varepsilon}{N}}}%
\title{Age of an allele and gene genealogies of nested subsamples for
  populations admitting large offspring numbers }
\author{Bjarki Eldon\\\\  Institut f\"ur Mathematik, Technische Universit\"at Berlin, \\
  10623 Berlin, Germany \\
  eldon@math.tu-berlin.de }
\date{\today}
\begin{document}%
\setlength{\baselineskip}{22pt}
\maketitle

Keywords: age of allele, multiple merger coalescent processes,
infinite alleles mutation model, Moran model, nested subsamples, weak
convergence

\textbf{Summary} Coalescent processes, including mutation, are derived
from Moran type population models admitting large offspring numbers.
Including mutation in the coalescent process allows for quantifying
the turnover of alleles by computing the distribution of the number of
original alleles still segregating in the population at a given time
in the past.  The turnover of alleles is considered for specific
classes of the Moran model admitting large offspring numbers.
Versions of the Kingman coalescent are also derived whose rates are
functions of the mean and variance of the offspring distribution.
High variance in the offspring distribution results in higher turnover
and younger age of alleles than predicted by the usual Kingman
coalescent.

\section{Introduction}

The age of an allele is central to understanding the nature of extant
variation.  The first mathematical treatment of the age of alleles
were done by \cite{kimura73}, \cite{maruyama74}, and
\cite{maruyama75}. In particular, \cite{kimura73} find, using
\cite{kimura64}'s diffusion model, that alleles in low frequency in
the population have surprisingly old expected ages.

More recently, coalescent techniques have been applied to the study of
the age of an allele
\citep{watterson84,saunders84,tavare84,donnelly86,griffiths98,stephens00}.
Two main extensions to the coalescent were obtained by
\cite{saunders84} One is the analysis of the ancestry of nested
subsamples, which, for example, yields the probability that a sample
shares its' most recent common ancestor with the whole population.
The other extension is the inclusion of neutral mutation in the
ancestral process, which, in turn, yields the probability that the
oldest allele in a population is included in a sample.  These
extensions are reviewed by \cite{tavare84}.  \cite{donnelly86}
consider age ordering of alleles in a sample, and obtain, for example,
the distribution of the number of representatives of the oldest allele
in a sample, a result originally due to \cite{kelly77}.  The ancestral
processes derived by \cite{watterson84}, \cite{saunders84}, 
\cite{tavare84} and \cite{donnelly86} were all based  on the usual
\cite{moran58:_random,moran62:statis} model of reproduction.  The
Moran model belongs to a large class of exchangeable population
models, originally introduced by \cite{cannings74}.  The Moran, and
indeed the celebrated Wright-Fisher \citep{fisher30,wright31} model
can be characterized as low offspring number models by allowing
individuals to have very many offspring with only negligible
probability.

Population models admitting large offspring numbers with
non-negligible probability
\citep{schweinsberg03,eldon06:coales,sargsyan08,huillet11} have
recently been proposed as appropriate for organisms with high
fecundity and high initial mortality, characteristics of many marine
taxa \citep{beckenbach94,hedgecock94does,hedgecock82,arnason04}, and
possibly forest trees \citep{ingvarsson10}. The gene genealogies of
samples drawn from populations admitting large offspring numbers are
characterised by multiple mergers of ancestral lineages
\citep{donnelly99,pitman99coales,sagitov99,schweinsberg00coales,mohle01},
or in which any number of active ancestral lineages may coalesce to
the same common ancestor.  In comparison, the Kingman coalescent
\citep{kingman82,kingman82b} allows at most two lineages to coalesce
each time.  One consequence of the multiple merger property is that
exact computations of many quantities of our interest can only be done
recursively.

The age and rate of turnover of alleles in a population admitting
large offspring numbers is our focus.  In the spirit of
\cite{saunders84} and \cite{tavare84}, we consider the coancestry of
nested subsamples drawn from a modified Moran model population, in
which the number of offspring contributed each timestep to the
population is random.  The rate of turnover of alleles associated with
the Kingman coalescent derived from populations with different
offspring distributions is considered, as well as probabilities of a
subsample carrying the oldest allele of a sample drawn from
populations admitting large offspring numbers.

\section{A population model of overlapping generations }

Consider a simple haploid population models of overlapping
generations, in which at each timestep a single individual (the
parent) chosen uniformly at random from the population contributes
offspring to replace those who perished in that timestep, so the
population size stays constant at $N$.  The coalescent processes
obtained from these population models will then be used to study the
coancestry process of nested subsamples.  Let $U$ denote the random
number of offspring produced by the parent, who always persists.  In
the usual Moran model, the parent always has one offspring ($U = 1$
a.s.).  \cite{eldon06:coales} consider a simple mixture distribution,
in which $U = 1$ with probability $1 - \varepsilon_{_N}$, and $U$
takes value $\lfloor \psi N \rfloor$ with probability
$\varepsilon_{_N}$, in which $\psi \in (0,1)$ is a constant.
\cite{eldon06:coales} use this simple model to illustrate the idea
that population models admitting large offspring numbers may be more
appropriate for high fecundity organisms than the usual Wright-Fisher
and Moran models.  A clear drawback of the model of
\cite{eldon06:coales} is that $\psi$ is a constant, which means that
the parent always has exactly the same number of parents, even when a
large offspring number event occurs.  Two simple extensions of the
usual Moran model will now be considered.  Our aim is to incorporate,
in a natural way, the randomness inherent in the number of offspring
contributed each timestep.

We will work with two models.  First, let us set notation.  Let $X$ be
a random variable with probability distribution $P_X$, in which
$\mathbb{P}[X \geq 1] = 1$, and finite moments, $\sup_N\mathbb{E}[X^k]
< \infty$ for all $k \geq 1$.  Let $Y$ be a random variable taking
values on the unit interval.  Finally, let $B$ be a Bernoulli random
variable with $\mathbb{P}[B = 1] = \varepsilon_{_N}$.  In model 1, we
take $U$ to be a simple mixture of $X$ and $Y$ (assuming $YN \in
\mathbb{N}$):
    \begin{equation}%
      \label{eq:defU}
      U = X(1 - B) + (NY)B
      \end{equation}%
      In model 2, we let $U$ be a Poisson random variable with random
      mean $M$, in which $M$ is a mixture of $X$ and $Y$, $M = X(1 -
      B) + (NY)B$.

      The idea behind the mixture distribution is the assumption that
      most of the time the parent contributes only few offspring to
      the population due to restrictions on resources such as space.
      Occasionally, however, there is disturbance in the environment,
      such as a storm or a forest fire.  The disturbance wipes away a
      fraction $Y$ of the population, allowing $NY$ offspring to take
      foothold in the population.  For simplicity, we assume that the
      difference between the number of individuals that perish, and
      the number of offspring replacing them, is negligible.  These
      ideas are similar to the ones proposed by \cite{schweinsberg03},
      who models the distribution of `potential' offspring, rather
      than the ones that actually take foothold in the population.
      However, Lemma~6 in \cite{schweinsberg03} establishes a link
      between the transition probability of the ancestral process and
      the distribution of the potential offspring.  The timescale in
      \cite{schweinsberg03}'s model depends on a parameter of the
      distribution of the potential offspring.  The timescale in our
      model will be independent of the distribution on $U$.  A similar
      idea of extending the usual Moran model is also raised by
      \cite{birkner09m}.  One can also view the mixture distribution
      as a simple model of a bottleneck followed by a rapid growth, or
      recolonization.  Indeed, \cite{grant98} suggest populations of
      sardines and anchovies may have been sucbject to periodic
      bottlenecks and recolonizations.

      Convergence of the coalescent process $\left(R_m; m \in
        \mathbb{N}_0\right)$ arising from both models is given in
      Appendix, as proof of convergence relies on standard theory.
      The coalescent process $\left(R_m;m \in \mathbb{N}\right) $
      without mutation describes the random ancestral relation among a
      set of $n$ individuals drawn at random from the population at
      time $m = 0$.  A state of the coalescent process $R_m$ is, by
      definition, the random equivalence relation containing
      individuals $(i,j)$ only if $i$ and $j$ share a common ancestor
      at timestep $r$ in the past.  One can think of the coalescent
      process as following labelled (or enumerated) ancestral lines
      back in time.

      Time in our model will be in units of $T_N \equiv N^2\wedge
      \varepsilon_{_N}^{-1}$, and will therefore be independent of any
      parameters associated with the offspring distribution on $X$ and
      $Y$.  Thus, if $N^2\varepsilon_{_N} \rightarrow 0$ then large
      offspring number events occur with negligible probability in a
      large population, and the coalescent process admits only mergers
      of two active ancestral lineages.  Otherwise the ancestral
      process admits multiple mergers of active ancestral lineages.

      For our purposes it suffices to consider the gene genealogical
      process $\left(A_t^{(\ell)}; t \ge 0 \right)$ that counts
      the number of ancestors, in which $A^{(\ell)}$ is
      associated with model $\ell \in \{1,2\}$.  To describe the rate matrix of
      $A^{(\ell)}$, write
        \begin{equation}%
          \label{eq:Bell}%
          \begin{split}%
          \beta_1 & = \mathbb{E}[X(X + 1)] \\
          \beta_2 & = 2\mathbb{E}[X] + \mathbb{E}[X^2] \\
          \end{split}%
          \end{equation}%
          If $N^2\sub{\varepsilon}{N} \rightarrow 0$ large offspring
          number events are negligible and the rate matrix $Q_\ell =
          \left(\sub{q}{{i,j}}^{(\ell)}\right)_{i,j \in [n]}$
          associated with model $\ell$ is given by
          \begin{equation}%
            \label{eq:QA1}%
            \sub{q}{{i,j}}^{(\ell)} = \begin{cases}%
            \binom{i}{2}\beta_\ell  & \textrm{if $j = i - 1$ } \\
            -\binom{i}{2}\beta_\ell & \textrm{if $j = i$} \\
            0 & \textrm{otherwise.}
            \end{cases}%
            \end{equation}%
            If $N^2\eN$ tends to a constant $\phi$ one obtains rate
            matrix $Q_\ell$ given by  
            \begin{equation}%
              \sub{q}{{i,j}}^{(\ell)} = \begin{cases}%
                \binom{i}{2}\left( \beta_\ell + \phi\mathbb{E}[Y^2(1 - Y)^{i - 2}] \right) & \textrm{if $j = i - 1$} \\
                \phi\binom{i}{k}\mathbb{E}[Y^k(1 - Y)^{i - k}] & \textrm{if $j = i - k + 1$}\\
                & 3 \le k \le i \\
                -\binom{i}{2}\beta_\ell -  \phi\mathbb{E}[1 - (1 - Y)^i - iY(1 - Y)^{i-1}]  & \textrm{if $j = i$} \\
                0 & \textrm{otherwise.}
                \end{cases}%
              \end{equation}%
              Finally, if $N^2\eN \rightarrow \infty$ large offspring
              number events are dominating and the rate matrix $Q$ has
              entries
            \begin{equation}%
              \label{eq:Qcase3}
               \sub{q}{{i,j}} = \begin{cases}%
                 \binom{i}{k}\mathbb{E}[Y^k(1 - Y)^{i - k}] & \textrm{if $j = i - k + 1$}\\
                & 2 \le k \le i \\
                -\left(1 - \mathbb{E}[(1 - Y)^i + iY(1 - Y)^{i - 1}] \right) & \textrm{if $j = i$} \\
                0 & \textrm{otherwise.}
                 \end{cases}%
              \end{equation}%
              In the present work we will primarily be concerned with
              three examples of a $\Lambda$ coalescent.  One is the
              process derived by \cite{schweinsberg03}, in which
              $\Lambda$ is the probability measure associated with the
              beta distribution, for $1 < \alpha < 2$,
              \begin{displaymath}%
                \Lambda(dx) =  \frac{1}{\Gamma(2 - \alpha)\Gamma(\alpha)}x^{1 - \alpha}(1 - x)^{\alpha - 1}dx.
                \end{displaymath}%
                In theory one can include the case $\alpha = 1$.  When
                $\alpha = 1$, however, the $\Lambda$ measure is simply
                the uniform distribution on $[0,1]$, which may not be
                biologically realistic.  For comparison, $\alpha$ is
                estimated to be around $1.5$ \citep{eldon11} for data
                on Atlantic cod \citep{arnason04}.  Another example of
                a $\Lambda$ coalescent we will consider is when
                $\Lambda$ is a scaled point mass at some point $\psi
                \in (0,1)$, $\Lambda(dx) = x^2\delta_\psi dx$.  The
                point mass process is the process derived by
                \cite{eldon06:coales}.  The third example we consider
                is when $Y$ takes the beta distribution with
                parameters $\alpha$ and $\beta$, in which case the
                rate $\sub{q}{{i,j}}$ for $j < i$ of the generator $Q$ in
                Equation~(\ref{eq:Qcase3}) takes  the form, with $2 \le k \le i$, 
                \begin{displaymath}%
                  \sub{q}{{i,j}} = \frac{B(k + \alpha, b - k + \alpha) }{B(\alpha, \beta)}
                  \end{displaymath}%
                  in which $B(\cdot,\cdot)$ is the beta function.
                  There is no particular biological reasoning behind
                  the choice of the beta distribution, other than $Y$
                  takes values between zero and one.  Allowing $Y$ to
                  be random seems a more realistic assumption than
                  requiring $Y$ to be fixed at some point between zero
                  and one.

              Turning to the rate matrix (\ref{eq:QA1}), it differs
              from the one associated with the usual Kingman
              coalescent by the factor $\beta_\ell$.  To compare our
              process to the Kingman coalescent, let $\mathbb{E}[T]$
              denote the expected value of the time to the most recent
              common ancestor associated with the Kingman coalescent,
              and with $\mathbb{E}[T^{(\ell)}]$ we denote the same
              quantity associated with model $\ell$.  Similarly, let
              $\mathbb{E}[S]$ denote the expected value of the total
              size of the gene genealogy associated with the Kingman
              coalescent, and $\mathbb{E}[S^{(\ell)}]$ the same
              quantity associated with model $\ell$.  One immediately
              obtains
              \begin{equation}%
                \label{eq:ETS}%
                \begin{split}
                \mathbb{E}[T^{(\ell)}] & = \frac{1}{\beta_\ell}\mathbb{E}[T] = \frac{2}{\beta_\ell}(1 - 1/n), \\  
                \mathbb{E}[S^{(\ell)}] & = \frac{1}{\beta_\ell}\mathbb{E}[S] = \frac{2}{\beta_\ell}  \sum_{i = 1 }^{n - 1}\frac{1}{i}, \quad \ell = 1,2. \\
                \end{split}%
                \end{equation}%
                Hence, populations with offspring distributions whose
                mean and variance are large are predicted to have
                lower genetic diversity, as measured in number of
                segregating sites, than populations with smaller mean
                and variance, if they have equal rates of mutation.
                The unit of time of a standard Moran model population,
                or of populations associated with either model 1 or 2
                is the same when $N^2\eN \rightarrow 0$, or $N^2$
                timesteps.  Thus, the results in
                Equation~(\ref{eq:ETS}) are directly comparable for
                populations with different offspring distribution.  To
                further compare the different population models, let
                $F_{S^{(\ell)}}$, $F_{T^{(\ell)}}$, $F_S$, and $F_T$
                denote the cumulative density functions of the random
                variables $S^{(\ell)}$, $T^{(\ell)}$, $S$, and $T$,
                respectively.  Elementary calculations now give
                $F_{S^{(\ell)}}(s) = F_{S}(\beta_\ell s)$, and
                $F_{T^{(\ell)}}(t) = F_{T}(\beta_\ell t)$ $(s,t \ge
                0)$. In words, the probability that $S^{(\ell)}$ and
                $T^{(\ell)}$ are no larger than some given values $s$
                and $t$, respectively, equals the probability that $S$
                and $T$ are no larger than $\beta_\ell s$ and $\beta_\ell t$,
                respectively.

                A further comparison between the usual Moran model,
                and models 1 and 2, can be made by considering the
                expected age $\mathbb{E}[\zeta_x]$ of a mutation with
                frequency $x \in (0,1)$.  In a Moran population, 
                \begin{equation}%
                  \label{eq:Ezeta}%
                  \mathbb{E}[\zeta_x] = \frac{-2x}{1- x}\log(x),
                  \end{equation}%
                  \citep{kimura73,griffiths98}.  The argument of
                  \cite{griffiths98} can be adapted to obtain an
                  expression of $\mathbb{E}[\zeta_x]$ in a population
                  with reproduction following models 1 or 2, to give
                   \begin{equation}%
                     \label{eq:Ezetal}%
                  \mathbb{E}[\zeta_x^{(\ell)}] = \frac{1}{\beta_\ell}\mathbb{E}[\zeta_x]
                  \end{equation}%
                  Indeed, let $\zeta_{n,b}$ denote the age of a
                  mutation in $b$ copies in a sample of $n$
                  individuals drawn from a usual Moran population, and
                  $\zeta_{n,b}^{(\ell)}$ the corresponding quantity
                  when drawn from a population with reproduction
                  following model 1 or 2.  Now,
                  \begin{displaymath}%
                    \mathbb{E}[\zeta_{n,b}^{(\ell)}] = \frac{1}{\beta_\ell} \mathbb{E}[\zeta_{n,b}],
                    \end{displaymath}%
                    in which $\mathbb{E}[\zeta_{n,b}]$ is given by
                    Equation~(5.4) in \cite{griffiths98}. Hence, by
                    taking appropriate limits as in
                    \cite{griffiths98}, one arrives at
                    Equation~(\ref{eq:Ezetal}).  Hence, the age of
                    mutations will tend to be overestimated if
                    Equation~(\ref{eq:Ezeta}) is applied to estimate
                    mutation age in a population with high mean and/or
                    variance in the offspring distribution.

\section{Turnover of ancestral lineages }

The question on the age of extant genetic variation can be addressed
using coalescent methods.   \cite{tavare84} and \cite{donnelly86}
consider the coalescent process including mutation, in which the
underlying reproduction mechanism is the usual Moran model.  Each time
a mutation event occurs, one lineage mutates into a `new' class, and
is hence no longer considered ancestral to the sample.  This is
equivalent of a lineage starting a new line of descent going forward
in time.

To study the turnover of ancestral lineages, we will work with the
lineage counting process $\left(\tilde{A}_r; r \in \mathbb{N}_0
\right)$ including mutation.  Since we allow mutation, the process
$\tilde{A}_r$ takes values in $\{n\} \equiv \{0, 1, \ldots, n\}$ if
the initial number of lineages is $n$.  When a mutation occurs on the
last of the original $n$ lineages, the process $\tilde{A}_r$ reaches
the absorbing state zero.  The way mutation is included in the
coalescent process is explained in Appendix.  Weak convergence of
$\tilde{A}_{\lfloor rT_N \rfloor}$ to a continuous-time process
$\tilde{A}_t$ follows from convergence of the coalescent
process $\tilde{R}$ including mutation (see Appendix).  If $N^2\eN \rightarrow 0$ the
rate matrix $\tilde{Q} \equiv \left(\sub{\tilde{q}}{{i,j}}\right)_{i,
  j \in \{ n\}}$ associated with process $\tilde{A}_t$ has entries
\begin{equation}%
  \label{eq:mQ1}%
  \sub{\tilde{q}}{{i,j}} = \begin{cases}%
    i\left(\theta + (i - 1)\beta_\ell\right)/2 & \textrm{if $j = i - 1$, $1 \le i \le n$} \\
    -  i\left(\theta + (i - 1)\beta_\ell\right)/2 & \textrm{if $j = i$, $0 \le i \le n$} \\
    \end{cases}%
  \end{equation}%
  in which $\beta_\ell$ was defined in Equation~(\ref{eq:Bell}).  If
  $N^2\eN \rightarrow \phi$ for some constant $\phi > 0$, the rate
  matrix  $\tilde{Q}$ has entries
  \begin{equation}%
    \sub{\tilde{q}}{{i,j}} = \begin{cases}%
      i((i - 1)\beta_\ell + \theta)/2 +  \phi\binom{i}{2}\mathbb{E}[Y^2(1 - Y)^{i - 2}] & \textrm{if $ j = i - 1$, $1 \le i \le n$ }\\
      \phi\binom{i}{k}\mathbb{E}[Y^k(1 - Y)^{i - k}] & \textrm{if $ j = i - k + 1$, $3 \le k \le n$ }\\
      -i((i - 1)\beta_\ell + \theta)/2 - \phi\mathbb{E}[1 - (1 - Y)^i - iY(1 - Y)^{i - 1}]  & \textrm{if $j = i$} \\
      0 & \textrm{otherwise} 
      \end{cases}%
    \end{equation}%
    Finally, if $N^2\eN \rightarrow \infty$ the entries of the rate
    matrix are
    \begin{equation}%
       \sub{\tilde{q}}{{i,j}} = \begin{cases}%
         i\theta/2  +  \binom{i}{2}\mathbb{E}[Y^2(1 - Y)^{i - 2}] & \textrm{if $j = i - 1$,  $1 \le i \le n$} \\
          \binom{i}{k}\mathbb{E}[Y^k(1 - Y)^{k - 2}] & \textrm{if $j = i - k + 1$, $3 \le k \le n$ } \\
          -i\theta/2 - \mathbb{E}[1 - (1 - Y)^i - iY(1 - Y)^{i - 1}]  & \textrm{if  $j = i$} \\
          0 & \textrm{otherwise.}%
         \end{cases}%
      \end{equation}%

      The rate of turnover of alleles is quantified with the
      probability $h_{i,j}(t) \equiv  \mathbb{P}[\tilde{A}_t = j | \tilde{A}_0 = i]$  given by   
      \begin{equation}%
        \label{eq:hij}%
        \begin{split}%
        h_{i,j}(t) &=  \sum_{k = j}^i e^{t\sub{\tilde{q}}{{k,k}}}r_{_i}^{(k)}\sub{\ell}{j}^{(k)}, \quad 1 \le j < i \\
        & = e^{t\sub{\tilde{q}}{{k,k}}}, \quad \quad\quad\quad\quad\quad j = i \\
        & = 1 - \sum\limits_{1 \le j \le i}h_{i,j}(t), \quad j = 0 \\
        \end{split}%
        \end{equation}%
        in which $r^{(k)}$ and $\ell^{(k)}$ are the right and left
        eigenvectors, respectively, of the rate matrix associated with
        $\tilde{A}$.  Only when the rate matrix of $\tilde{A}$ is of
        the form (\ref{eq:mQ1}), a closed form expression for the
        eigenvectors can be obtained as follows.  Define, for some
        constant $c > 0$, the factorials
  \begin{equation}%
    \label{eq:cfactors}%
    \begin{split}
    a_{(c,k)} &\equiv  a(a + c)(a + 2c)\cdots(a + (k-1)c),\\
    a_{[c,k]} &\equiv  a(a - c)(a - 2c)\cdots(a - (k-1)c).
    \end{split}%
    \end{equation}%

    The results of \cite{tavare84} can be adapted to obtain the left
    $\ell^{(k)}$ and right $r^{(k)}$ eigenvectors of the transition
    matrix given by Equation~(\ref{eq:mQ1}).  One obtaines
    $\ell_j^{(0)} = \delta_{j,0}$, and
  \begin{equation}%
    \label{eq:leigenvc}%
    \begin{matrix}
      \ell_j^{(k)} & = & 0, &  \quad j > k \ge 1; \\
      \ell_j^{(k)} & = & \dbinom{k}{j}(-1)^{k - j}\dfrac{(cj + \theta)_{(c,k - 1)} }{(ck + \theta)_{(c,k - 1)}  }, & j \le k \\
      \end{matrix}%
    \end{equation}%
    in which $c = B_\ell$.  In the same way, $r_j^{(0)} = 1$ for all
    $j$, and
\begin{equation}%
  \label{eq:reigenvc}%
  \begin{matrix}%
  r_j^{(k)} & = & 0, & j < k \\
  r_j^{(k)} & = & \dbinom{j}{k}\dfrac{(ck + \theta)_{(c,k)}}{(cj + \theta)_{(c,k)} }, & j \ge k.
  \end{matrix}%
  \end{equation}%
  Equations~(\ref{eq:hij}--\ref{eq:reigenvc}) now give, with $i_{[k]} \equiv i(i-1)\cdots(i - k + 1)$, 
  \begin{equation}%
    \label{eq:hijt}%
    h_{i,j}(t) = \sum_{k = j}^i e^{t\sub{\tilde{q}}{{k,k}}}(-1)^{k - j}\frac{i_{[k]}}{j!(k - j)!}\left( (2k - 1)c + \theta \right) \frac{(cj + \theta)_{(c,k-1)} }{(ci + \theta)_{(c,k)} }, \quad 1 \le j < i.
    \end{equation}%

  One can also view the constant $c$ as a population size scaling
  constant.  The probability $h_j(t)$ that $j$ of original lines of
  descent from the whole population are still present at time $t$ is
  given by
    \begin{equation}%
      \label{eq:hjt}%
      \begin{split}%
      h_j(t) & = \sum_{k=j}^\infty e^{t\sub{\tilde{q}}{{k,k}}}(-1)^{k - j}\left( (2k - 1)c + \theta \right)  \frac{(cj + \theta)_{(c,k - 1)} }{c^k j!(k - j)! }, \quad 1 \le j;  \\
      h_0(t) & = 1 + \sum_{k= 1}^\infty e^{t\sub{\tilde{q}}{{k,k}}}(-1)^{k}( (2k - 1)c + \theta)\frac{\theta_{(c,k-1)} }{c^k k! },  \\
      \end{split}%
      \end{equation}%
      obtained by taking $i \rightarrow \infty$.  The formulas for
      $h_{i,j}(t)$ and $h_{j}(t)$ associated with the usual Kingman
      coalescent with mutation are recovered by taking $c = 1$.  Even
      though the sum in Equation~(\ref{eq:hjt}) is infinite, the terms
      quickly become small.  Even moderate values of $\theta$ and $c$
      $(= \beta_\ell)$ lead to quick turnover of alleles in the population
      (Table~1).

      The distribution of the number of ancestral lineages of either a
      finite sample, or, of the whole population, at any given time,
      is straightforward to obtain using the results of
      \cite{littler75}, \cite{griffiths79}, \cite{griffiths80},
      \cite{watterson82}, \cite{kingman82b}, and \cite{tavare84}.  The
      form of the eigenvectors associated with the usual Kingman
      coalescent is the same as when associated with the modified
      coalescent derived from models 1 or 2; only the eigenvalues of the
      corresponding rate matrices differ in an obvious way.

      An explicit form of $h_j(t)$ is hard to obtain when associated
      with a $\Lambda$ coalescent since the rate matrix is triagonal,
      and closed form expressions for the eigenvectors are not easy to
      obtain.  The eigenvectors can, however, be computed recursively,
      as shown in Appendix.  The probability $h_{i0}(t)$ that all of
      $i$ initial ancestral lineages have vanished by time $t$ is
      graphed as a function of time $t$ in Figure~1 for two examples
      of a $\Lambda$ coalescent.  In Figure~1a is the point mass
      process $\Lambda(d\psi) = \psi^2\delta_\psi d\psi$ studied by
      \cite{eldon06:coales}, and in Figure~1b is the beta coalescent
      derived by \cite{schweinsberg03}.  In the beta coalescent, time
      is in units proportional to $N^{\alpha - 1}$.  Hence, the
      results in Figure~1b must be interpreted with the timescale
      property of $\alpha$ in mind.  The $\psi$ parameter of the point
      mass process is not a timescale parameter; the values of
      $h_{i0}(t)$ for different values of $\psi$ can therefore be
      directly compared.  The quantity $h_{i0}(t)$ can also be
      interpreted as the probability that the oldest mutation in a
      sample of $i$ sequences is not older than $t$ time units.
      Figure~1c reports values of $h_{i0}(t)$ when the random variable
      $Y$ has the beta distribution with parameters $(\alpha,\beta)$.
      The values of $h_{i0}(t)$ in Figure~1c are directly comparable
      for different values of the parameters $\alpha$ and $\beta$
      since they are not timescale parameters.  If $\theta$ is large,
      ancestral lineages will vanish quickly from a population
      admitting large offspring numbers, one concludes from Figure~1.
      In addition, the effects of $\theta$ seem stronger than those of
      the coalescence parameters, for the range of parameter values
      considered in Figure~1.

\section{The subsample process}

Two alleles drawn at random from a standard Moran population have
probability $1/3$ of sharing their most recent common ancestor with
the whole population present at the time the two alleles were sampled.
This result follows from the probability $(j-1)(i+1)/[(j+1)(i - 1)]$
that $j$ of $i$ lineages share a most recent common ancestor with the
$i$ lineages \citep{saunders84}.  This result indicates that even
samples of even moderate size will yield a good estimate of the time
to the most recent common ancestor of the whole population.  To
further study the rate of turnover of alleles in populations admitting
large offspring numbers, we consider simple properties of the gene
genealogical process of nested subsamples.

One can recursively compute the conditional distribution of the number
of ancestors of the subsample, given the number of ancestors of the
sample, as shown in section \ref{sec:subsample}.  Closed-form
expressions can be obtained for a population with the Kingman
coalescent \citep{littler75,griffiths80,kingman82b,tavare84}.  The
presence of multiple mergers makes closed-form expressions hard to
obtain for a general $\Lambda$ coalescent.  In this section we will be
concerned with the probability that a subsample shares it's most
recent common ancestor with the sample.  Denote by $A_1^*(u)$ the number
of ancestors of the sample when the the number of ancestors of the
subsample first reaches a given value $u$.  Similary, denote by $A_2^*(v)$
the number of ancestors of the subsample when the number $A_1$ of
ancestors of the sample first reaches some value $v$.  Closed-form
expressions for the distributions of $A_1^*$ and $A_2^*$ have been
obtained for a usual Moran population \citep{saunders84}.  The
conditional distributions of $A_1^*$ and $A_2^*$ when associated with
a $\Lambda$ coalescent may be computed recursively, as shown in
Appendix.

To compute our quantity of interest, let $\psi_{i,j}(u,v) \equiv
\mathbb{P}[A_1^*(v) = u | A_1(0) = i, A_2(0) = j]$.  The probability
$\psi_{i,j}(1,1)$ that the subsample shares its' most recent common
ancestor with the whole sample is straightforward to compute
recursively as follows. Writing $i^\prime = i - k + 1$, $j^\prime = (j -
\ell + 1)1_{\{ \ell > 0 \}} + j1_{\{\ell = 0\}}$, one obtains
  \begin{equation}%
    \label{eq:psiij11}
    \psi_{i,j}(1,1) = \sum_{k = 2}^{i}\beta(i,i^\prime) \sum_{\ell = 0}^{j\wedge k}\frac{\binom{i-j}{k - \ell}\binom{j}{\ell} }{\binom{i}{k}}\psi_{i^\prime, j^\prime}(1,1)
    \end{equation}%
    with the boundary conditions $\psi_{j,j}(1,1) = 1$ and
    $\psi_{i,1}(1,1) = 0$ for $i > 1$.  The result obtained from the
    usual Moran model, $ \psi_{i,j}(1,1) = (j-1)(i + 1)/[(j+1)(i -
    1)]$, can be recovered from Equation~(\ref{eq:psiij11}).  The
    recursion (\ref{eq:psiij11}) simplifies considerably when $j =
    2$. Writing $\psi(i) = \psi_{i,2}(1,1)$, one obtains
    \begin{equation}%
      \label{eq:psii2}
      \psi(i) =  \sum_{k = 2}^{i-2} \beta(i,i - k + 1)\frac{(i - k)(i + k - 1)}{i(i - 1)}\psi(i - k + 1) + \beta(i, 2)\frac{2}{i} + \beta(i,1).
      \end{equation}%
      When associated with a $\Lambda$ coalescent, $\psi(i)$ can be
      quite small (Table~\ref{tab:table2}).  For comparison, the
      corresponding values when associated with the Kingman coalescent
      are $\psi(500) = 0.335$, and $\psi(1000) = 0.334$.  An indicator
      of how well a sample represents the population it is drawn from
      may be taken as the probability $\psi_{\infty,j}(1,1)$.  A
      sample of moderate size drawn from a standard Moran population
      represents the population quite well, since then
      $\psi_{\infty,j}(1,1) = (j - 1)/(j + 1)$.  The results in Table~(\ref{tab:table2})
      indicate that only a large sample drawn from a population
      admitting large offspring numbers has a good chance of sharing
      its' most recent common ancestor with the whole population,  if
      we interpret the sample as the whole population.

      The conditional distributions of $A_1^*$ and $A_2^*$ will be the
      same for models 1 and 2 as for the usual Moran model.  Hence,
      samples of size $j$ drawn from populations following a large class of
      reproduction law, or from populations differing in their
      population size, will all have probability $(j -1)/(j + 1)$ of
      sharing their most recent common ancestor with the population
      present at the time of sampling.

\section{The subsample process including mutation}

The age of an allele is fundamental to population genetics.  A sample of
size $j$ has probability $j/(j + \theta)$ of including the oldest
allele in the population, if drawn from a usual Moran population
\citep{watterson77,saunders84,tavare84}.  \cite{slatkin00} bases a
test for selection on allele age.   Results concerning age of
alleles can be obtained by considering the ancestral process of nested
subsamples when the ancestral process includes mutation.

One might wish to know how the expected number of ancestors of a
sample changes given the number of ancestors of a subsample.  Define
$\tilde{A}_1^*$ and $\tilde{A}_2^*$ in the same way as $A_1^*(a)$ and
$A_2^*(a)$, respectively.  The quantity $\tilde{A}_1^*(u)$ denotes the
number of ancestors of the sample, including mutation, when the number
of ancestors of the subsample first takes value $u$.  The conditional
distributions of $\tilde{A}_1^*$ and $\tilde{A}_2^*$ can be computed
recursively (see Appendix) when associated with a population admitting
large offspring numbers.  The conditional expected value
$\mathbb{E}[\tilde{A}_2^*(\sub{a}{1})]$ is graphed as a function of
the number $\sub{a}{1}$ of ancestors of the sample for  three
examples of a $\Lambda$ coalescent.  The idea behind calculating
$\mathbb{E}[\tilde{A}_2^*(\sub{a}{1})]$ is to infer how the number of
ancestors of a sample changes relative to the whole population, by
treating the sample as the population.  The number of ancestors of a
sample drawn from a population admitting large offspring numbers would
then, on average, be expected to be smaller than if drawn from a usual
Moran population, for a given number of ancestors for the whole
population.  Furthermore, mutation rate appears to matter little for
the Kingman coalescent, while being a more important player for the
$\Lambda$ coalescent examples we are considering (Figure~2).  The results in
Figure~2 suggest that the expected number
$\mathbb{E}_\Lambda[\tilde{A}_2^*]$ of ancestors of the subsample when
associated with a $\Lambda$ coalescent would tend to be smaller than
the corresponding value $\mathbb{E}_K[\tilde{A}_2^*]$ associated with
the Kingman coalescent.  In addition,
$\mathbb{E}_\Lambda[\tilde{A}_2^*]$ may be convex in some cases, while
$\mathbb{E}_K[\tilde{A}_2^*]$ is a concave function of the number of
ancestors of the sample.  In Figure~3, values of the number
$\mathbb{E}_\Lambda[\tilde{A}_1^*]$ of ancestors of the sample when
associated with a $\Lambda$ coalescent are graphed as a function of
the number of ancestors of the subsample.  The results in Figure~3
show that $\mathbb{E}_\Lambda[\tilde{A}_1^*]$ may be either smaller or
larger than the corresponding value $\mathbb{E}_K[\tilde{A}_1^*]$
associated with the Kingman coalescent.  In addition, while
$\mathbb{E}_K[\tilde{A}_1^*]$ is a convex function of the number of
ancestors of the subsample, $\mathbb{E}_\Lambda[\tilde{A}_1^*]$ seems
to range from being convex to appearing an almost linear function of
the number of ancestors of the subsample.

An expression for the probability that a subsample of a sample drawn
from a usual Moran model population carries the oldest allele of the
sample yields a remarkably simple expression for the probability that
the sample carries the oldest allele of the population.  The
probability $\tilde{u}(i,j) \equiv \tilde{\psi}_{i,j}(0,0)$ that a
subsample includes the oldest allele of a sample drawn from a
population admitting large offspring numbers is straightforward to
compute recursively, namely
\begin{equation}%
  \label{eq:muij}%
  \begin{split}%
  \tilde{u}(i,j) & = \frac{j\theta/2}{\lambda(i)}\tilde{u}(i - 1, j - 1) + \frac{(i - j)\theta/2}{\lambda(i)}\tilde{u}(i - 1,j) \\\\
  & \quad + \sum_{k=1}^{i- 1}\beta(i, i - k) \sum_{\ell = 0}^{k\wedge j}\frac{\binom{j}{\ell }\binom{i - j }{k - \ell} }{\binom{i}{k + 1} } \tilde{u}(i - k, j - \ell1_{\{ \ell > 1 \} })\\
  \end{split}%
  \end{equation}%
  with boundary probabilities $\tilde{u}(j,j) = 1$, and  $\tilde{u}(i,0) =
  0$ if $i > 0$.  Table~3 shows values of $\tilde{u}(1000,2)$ varying
  over $\theta$ and $\pi$.  The corresponding values for the Kingman
  coalescent are $\tilde{u}(1000,2) = 0.667$ when $\theta = 1$, and
  $\tilde{u}(1000,2) = 0.287$ when $\theta = 5$
  \citep{kelly77,saunders84}.  The results in Table~3 indicate that
  two lineages drawn from a population admitting large offspring
  numbers will, in some cases, have quite a low probability of being
  the oldest allele of the population, if we interpret the sample as
  the population.  One might also be interested in how quickly the
  probability of the subsample carrying the oldest allele of the
  sample increases with the subsample size.  Figure~4 reports values
  of $\tilde{u}(150,j)$ as a function of $j$ for the three example
  $\Lambda$ coalescents.  While $\tilde{u}(i,j)$ is a concave function
  of $j$, and thus increases quickly with $j$ when associated with the
  Kingman coalescent, $\tilde{u}(i,j)$ increases slower, or almost
  linearly in some cases, when associated with a $\Lambda$ coalescent.
  Thus, a subsample of a sample drawn from a population admitting
  large offspring numbers has smaller probability of carrying the
  oldest allele of the sample than if the sample was drawn from a
  usual Moran population.

  The conditional distributions of $\tilde{A}_1^*$ and $\tilde{A}_2^*$
  derived from models 1 or 2 will be different from the distributions
  derived from the usual Moran model.  Denote the conditional
  distribution of $\tilde{A}_2^*$ by $\phi(\ell_1, \ell_2)$,
  \begin{displaymath}
\phi(\ell_1,  \ell_2) \equiv \mathbb{P}(\tilde{A}_1^*(\ell_1) = \ell_2 | \tilde{A}_1(0) = i, \tilde{A}_2(0) = j ).  
\end{displaymath}%
\cite{saunders84} obtain  $\phi(\ell_1, \ell_2)$  by solving the recursion (Theorem~6 in \citep{saunders84})
\begin{equation}%
  \phi(\ell_1, \ell_2) = \phi(\ell_1 + 1, \ell_2)\frac{(\ell_1 - \ell_2 + 1)(\ell_1 + \ell_2 + \theta) }{(\ell_1 + 1)(\ell_1 + \theta) }  +  \phi(\ell_1 + 1, \ell_2 + 1)\frac{(\ell_2 + 1)(\ell_2 + \theta) }{(\ell_1 + 1)(\ell_1 + \theta) }
  \end{equation}%
  with $\phi(i,\ell_2) = 1$ if $\ell_2 = j$, and $\phi(i,\ell_2) = 0$
  otherwise.  In a population following model 1 or 2,   one obtains    
  \begin{equation}%
    \begin{split}
  \phi(\ell_1, \ell_2) & = \phi(\ell_1 + 1, \ell_2)\frac{(\ell_1 - \ell_2 + 1)((\ell_1 + \ell_2)\beta + \theta) }{(\ell_1 + 1)(\ell_1\beta + \theta) }  +   \phi(\ell_1 + 1, \ell_2 + 1)\frac{(\ell_2 + 1)(\ell_2\beta + \theta) }{(\ell_1 + 1)(\ell_1\beta + \theta) } \\\\
  & = \phi(\ell_1 + 1, \ell_2)\frac{(\ell_1 - \ell_2 + 1)(\ell_1 + \ell_2 + \theta^*) }{(\ell_1 + 1)(\ell_1 + \theta^*) }  +   \phi(\ell_1 + 1, \ell_2 + 1)\frac{(\ell_2 + 1)(\ell_2 + \theta^*) }{(\ell_1 + 1)(\ell_1 + \theta^*) }
  \end{split}
  \end{equation}%
  where we subpress the subscript $\ell$ on $\beta$ $(= \beta_\ell)$,
  and $\theta^* = \theta/\beta$.  Hence, the conditional distribution
  of $\tilde{A}_2^*$ is given by Theorem~6 in \cite{saunders84}, with
  $\theta$ replaced by $\theta/\beta$.  The same applies to the
  conditional distribution of $\tilde{A}_1^*$ (see Theorem~7 in
  \cite{saunders84}).  In this context we must mention the scaling of
  mutation. We insist that mutation scale in units of $N^2$ timesteps
  exactly, in order for the models to be comparable.  Hence, the
  mutation rate $\theta$ is the same quantity for all models that
  yield a Kingman like coalescent.  Thus, the probability that the
  oldest allele of $i$ lineages is among a subset of $j$ lineages is
  $j(i\beta + \theta)/[i(j\beta + \theta)]$, when associated with
  model 1 or 2.  The oldest allele of a population associated with
  model 1 or 2 will therefore be among $j$ lineages with probability
  $j\beta/(j\beta + \theta)$.  The constant $\beta$ can also be
  interpreted as a population size scaling constant, if we assume that
  time is scaled in units of $N^2$ timesteps.  If the size of a Moran
  population is scaled by factor $c$,  then $\beta = 1/c^2$.

  \cite{kelly77} and \cite{saunders84} obtain the distribution of the
  number of representatives $F_i$ of the oldest allele in the sample
  of size $i$ drawn from a population with the usual Moran
  reproduction.  Using the same argument as in \cite{saunders84}, one
  can recursively compute the distribution of $F_i$, using
  Equation~(\ref{eq:muij}).  The expected value $\mathbb{E}[F_i]$ of
  $F_i$ when associated with a $\Lambda$ coalescent tends to be
  smaller than when associated with the Kingman coalescent (Table~4).
  The magnitude of the difference also varies with the type of the
  $\Lambda$ coalescent, with values of $\mathbb{E}[F_i]$ associated
  with the beta coalescent being quite similar to the ones associated
  with the Kingman coalescent.  In contrast, $\mathbb{E}[F_i]$ seems
  to converge to one as $\psi$ decreases.  The expected value of $F_i$
  also tends to one when $\theta$ increases, when associated with the
  usual Kingman coalescent.  When associated with model 1 or 2, and
  large offspring number events are negligible, the results of
  \cite{kelly77} and \cite{saunders84} yield
  \begin{displaymath}%
    \mathbb{E}[F_i] = \frac{i\beta + \theta}{\beta + \theta}, \quad \textrm{Var}[F_i] = \frac{\beta(\theta + \beta)\theta(i - 1) }{(\beta + \theta)^2(2\beta + \theta) }.
    \end{displaymath}%

\section{Discussion and Conclusion}

The Wright-Fisher and Moran models have been the basis of the majority
of work in theoretical population genetics.  However, they make strong
assumptions about the offspring distribution of individuals, that are
not fulfilled by most natural populations.  The simple generalized
versions of the Moran model developed in the present work allow one,
in a natural way, to account for at least some of the stochasticity
inherent in natural populations.  A further benefit is that the
modified population models run on the same timescale, and are thus
directly comparable.

Population models admitting large offspring numbers are natural
candidates for populations with highly fecund individuals, including a
diverse group of marine taxa.  In this work we consider a model in
which a single individual contributes a random number of offspring at
each timestep.  An important open question is the distribution of the
number of offspring, which will require insight into the biology and
ecology of natural populations.  Accurately modeling offspring
distribution may be particularly important in conservation genetics
\citep{amos01}.  By incorporating the mean and variance of the
offspring distribution into the coalescence rates, we show that
populations with high variation in their offspring distribution will
have lower genetic diversity than predicted by the usual Moran model,
as is often observed in particular among marine populations.
Admitting large offspring numbers leads to multiple merger coalescent
processes, which can predict the starlike pattern of haplotypes often
observed among marine fishes \citep{grant98}.

The multiple merger coalescent processes derived from population
models admitting large offspring numbers are more difficult to handle
mathematically than any version of a Kingman coalescent.  Closed-form
analytic expressions are hard to obtain. Exact computations of many
quantitites can therefore only be done recursively when associated
with multiple merger processes.  Recursive computations limit us in
some cases to small sample sizes.  Important insights can still be
obtained, and which can give clues to results for large sample sizes.

In this work we consider the rate of turnover of alleles in
populations while paying attention to the offspring distribution.  Our
main conclusion is that the rate of turnover of alleles increases with
the variance in the offspring distribution.  This means that alleles
will tend to be younger than in the usual Wright-Fisher or Moran
populations.  In addition, our results suggest that one may need large
sample sizes to accurately predict the age of the most recent common
ancestor of a population in populations with large offspring variance.
In addition, one may need to draw a large sample from some populations
admitting large offspring numbers before having a good chance of
having caught the oldest allele of the population.

{\small The author was supported in part by EPSRC grant EP/G052026/1,
  in part by DFG grant BL 1105/3-1, and by a Junior Research
  Fellowship at Lady Margaret Hall, University of Oxford.  }

\bibliographystyle{/home/markov/eldon/verk/vinna/vinna0/spmpscinat}%
\bibliography{epsrc}

\section{Appendix}%

\subsection{Weak convergence of the coalescent process including mutation}

In this section we establish weak convergence in the function space
$\sub{D}{\tilde{E}}([0,\infty))$ of the coalescent process including
mutation and derived from a general Cannings \cite{cannings74} model
admitting large offspring numbers.  In a Cannings model of a
population of constant size $N$, the offspring variables
$\sub{\nu}{i}$ are exchangeable, and $\sub{\nu}{1} + \cdots +
\sub{\nu}{N} = N$.  By convention, let $\sub{D}{E}([0,\infty))$ denote
the space of right continuous functions with left limits on
$[0,\infty)$ with values in $E$, the set of all equivalence relations
on $\{1, \ldots, n \}$.  The coalescent process $\left(R_m\right)_{m
  \in \mathbb{N}_0}$ without mutation describes the random ancestral
relation among a set of $n$ individuals drawn at random from the
population at time $m = 0$.  A state $\xi$ of the process
$\left(R_m\right)_{m \in \mathbb{N}_0}$ is, by definition, the random
equivalence relation containing individuals $(i,j)$ only if $i$ and
$j$ share a common ancestor at timestep $r$ in the past.  The
coalescent process starts in state $\left\{(i,i): 1 \le i \le n
\right\}$. The absorbing state is $ \left\{ (i,j): 1 \le i,j \le n
\right\}$.  Let $b \equiv b_1 + \cdots + b_a$, in which $b_1, \ldots,
b_a$ denote the number of classes of $\xi$ that merge into each of $a$
classes of $\eta$.  Write $(m)_n \equiv m(m - 1)\cdots (m - n + 1)$,
$(m)_0 \equiv 1$.  Under an exchangeable \cite{cannings74}
reproduction model and fixed population size $N$, the transition
probability $ P_{\xi,\eta}(N) = \mathbb{P}[R_{m + 1} = \eta | R_m =
\xi ]$ from state $\xi$ to $\eta$ is given by
\begin{equation}%
  P_{\xi,\eta}(N) = \dfrac{(N)_a}{(N)_{b}}\mathbb{E}\left[ \left(\sub{\nu}{1}\right)_{b_1}\cdots\left( \sub{\nu}{a} \right)_{b_a} \right]
  \end{equation}%
  Time is scaled in units of $1/\sub{c}{N}$, in which $\sub{c}{N}
  \equiv \mathbb{E}[\sub{\nu}{1}(\sub{\nu}{1} - 1)]/(N-1)$. To pass to a
  continuous-time process, we require that $\lim_{N \rightarrow
    \infty}\sub{c}{N} = 0$.

  We will now list the conditions on the population model under which
  the scaled process $\left(R_{\lfloor t/\sub{c}{N} \rfloor
    }^{(N)}\right)_{t \ge 0}$ converges weakly to
  $\left(R_t\right)_{t \ge 0}$ \citep{sagitov99}. Denote by
  $\Lambda$  a finite measure on the Borel subsets of the unit
  interval. The three conditions are
  \begin{equation}%
    \label{eq:conditions}%
    \begin{split}%
      \mathbb{E}[\left(\sub{\nu}{1} - 1\right)^2] & = o(N);  \\
      \lim_{N \rightarrow \infty} \frac{N}{\sub{c}{N}}\mathbb{P}[\sub{\nu}{1} > Nx] &= \int_x^1 y^{-2}\Lambda(dy),\quad 0 < x < 1; \\
      \lim_{N \rightarrow \infty}\frac{\mathbb{E}[(\sub{\nu}{1} - 1)^2\cdots (\sub{\nu}{a} - 1)^2] }{\sub{c}{N}N^a } & = 0, \quad a \ge 2;\\
      \end{split}%
    \end{equation}%
       \citep{sagitov99}.   Now define 
  \begin{equation}%
    p_{_{\xi,\eta}}(N) = \dfrac{(N)_a}{(N)_{b}}\mathbb{E}\left[  \left(\sub{\nu}{1}\right)_{b_1}\cdots\left( \sub{\nu}{a} \right)_{b_a}  \right], \quad b_1 \geq 2, \quad b_2 = \cdots = b_a = 1.
    \end{equation}%
   The conditions in Equation~(\ref{eq:conditions}) yield
   \begin{equation}%
    \label{eq:plimit}%
   \frac{1}{c_{_N}} p_{{_\xi},{_\eta}}(N)   \rightarrow \int_0^1x^{k-2}(1-x)^{b-k}\Lambda(dx), \quad k \geq 2, \quad a \geq 1 
    \end{equation}%
    \citep{sagitov99}, in which $b_1 = k$, $b_2 = \cdots = b_a = 1$.

  The coalescent process $\left(\tilde{R}_m\right)_{m \in
    \mathbb{N}}$ with mutation separates each relation into mutant
  (`new') and non-mutant (`old') equivalence classes.  The terminology
  `old' and `new' is to remind that mutations are always to new types
  (alleles), and hence start new lines of descent
  \citep{tavare84,donnelly86}.  Only non-mutant equivalence classes
  either coalesce or mutate.  Let $\tilde{E}$ denote the set of all
  equivalence classes on $\{1, \ldots, n\}$ in which each class is
  partitioned into old and new classes (see \cite{mohle99weak}).  The
  total number of old classes of the current state $\xi$ is $b + m$.
  The transition probability for the transition from $\xi$ to $\eta$
  is
\begin{equation}%
  P_{\xi,\eta}(N) = \mu_{_N}^m(1 - \mu_{_N})^{b}\dfrac{(N)_a}{(N)_{b}}\mathbb{E}\left[  \left(\sub{\nu}{1}\right)_{b_1}\cdots\left( \sub{\nu}{a} \right)_{b_a}  \right] 
  \end{equation}%
  To retain mutations in the limit process, we require that the
  mutation probability $\mu_{_N}$ scales according to $c_{_N}$, namely
     \begin{equation}%
       \label{eq:mulimit}%
     \lim_{N \rightarrow \infty}  \dfrac{\mu_{_N}}{c_{_N}}  = \dfrac{\theta}{2}, \quad 0 < \theta < \infty.
       \end{equation}%
       in which $\theta$ is a constant. By $\xi \rightsquigarrow \eta$
       denote a transition due to mutation, in which one old class of
       $\xi$ becomes a new class of $\eta$.  By $\xi \prec \eta$
       denote a transition from $\xi$ to $\eta$ in which $\xi \subset
       \eta$ and $\eta$ denotes a merger of at least two of the
       equivalence classes of $\xi$.  Define a sequence
       $\sub{\tilde{Q}}{N} = \left( \sub{\tilde{q}}{N}(\xi,\eta)
       \right)_{\eta,\xi \in \tilde{E}}$ of generators with
       entries \begin{equation}%
     \label{eq:qNdef}
     \tilde{q}_{_N}(\xi,\eta) = \begin{cases}   p_{_{\xi,\eta}}(N)/\sub{c}{N} & \textrm{if $\xi \prec \eta$} \\\\
       \frac{\mu_{_N}}{\sub{c}{N}}\mathbb{E}\left[\sub{\nu}{1}\cdots \sub{\nu}{a}\right] & \textrm{if $\xi \rightsquigarrow \eta $} \\\\ 
        -\sum\limits_{\substack{ \xi \curlyeqprec \eta \\ \xi \rightsquigarrow \eta}}q_{_{\xi,\eta}}^{(N)} & \textrm{if $\xi = \eta$} \\\\
       0 & \textrm{otherwise}.
       \end{cases}
     \end{equation}%
     Equations (\ref{eq:plimit}--\ref{eq:qNdef}) now imply that
     the sequence $\sub{\tilde{Q}}{N}$ of generators converges
     entry-wise to a generator $\tilde{Q} = \left( \tilde{q}(\xi,\eta)
     \right)_{\eta,\xi \in\tilde{E}}$ with entries
       \begin{equation}%
  \label{eq:qNcN}%
  \tilde{q}(\xi,\eta) =  \begin{cases}
     \int_{0}^1x^{k - 2}(1 - x)^{|\xi| - k}\Lambda(dx)  & \textrm{if $\xi \prec \eta$}\\\\
      \theta/2 & \textrm{if  $\xi \rightsquigarrow \eta $} \\\\
      -|\xi|\frac{\theta}{2} - \int_0^1(1 - (1 - x)^{|\xi|} - |\xi|x(1 - x)^{|\xi| - 1})x^{-2}\Lambda(dx)  & \textrm{if  $\eta = \xi $} \\\\
       0 & \textrm{otherwise}.
       \end{cases}
  \end{equation}%
  Since the state space $\tilde{E}$ is finite, the entry-wise convergence
  $\sub{\tilde{Q}}{N} \rightarrow \tilde{Q}$ implies weak convergence
  in $\sub{D}{\tilde{E}}([0,\infty))$ of $\tilde{R}_{\lfloor
    t/\sub{c}{N} \rfloor }$ to   $\left(\tilde{R}_t; t \ge 0
  \right)$ with generator $\tilde{Q}$, by Theorem 17.25(i) in
  \cite{kallenberg97}.

\subsection{Generators of the coalescent processes}

In this section we describe the generators of the coalescent processes
obtained from  models 1 and 2.   
    
By $E$ denote the set of all equivalence relations on $\{1, \ldots,
n\}$.  One readily obtains convergence of the finite-dimensional
distributions of $R_m$.  To prove convergence of the
time-scaled process $R_{\lfloor tT_N \rfloor}$ in the space
$D_E[0,\infty)$ of all functions on $[0,\infty)$ with values in $E$
that are right continuous and with left limits, we apply the
convergence result of the previous section and obtain that the
ancestral process $\left( R_{\lfloor tT_N \rfloor}; t \ge
  0\right)$ when associated with population model 1 or 2 converges
weakly in $D_E[0,\infty)$ to a continuous-time process
$\left(R_t; t \ge 0 \right)$ whose rate matrix $Q$ depends
on the limit $N^2\varepsilon_{_N}$.  To describe the transition
matrices, define $\sub{\alpha}{\ell}$ associated with model $\ell \in \{1,2\}$, 
    \begin{equation}%
      \label{eq:alphadef}%
      \begin{split}%
    \sub{\alpha}{1} &\equiv \mathbb{E}[X^2] \\
    \sub{\alpha}{2} & \equiv \mathbb{E}[X(X + 1)]
    \end{split}
    \end{equation}%
    The constants $\sub{\alpha}{\ell}$ play similar role as the
    constants $\beta_\ell$ associated with the gene genelaogical
    processes.  Let $|\xi|$ denote the number of classes (lineages) in
    relation $\xi$.  If $N^2\varepsilon_{_N} \rightarrow 0$ the rate
    matrix $Q_\ell = \left(q_{_{\xi,\eta}}^{(\ell)}\right)_{\xi,\eta
      \in E}$ associated with model $\ell \in \{1,2\}$ is given by
    \begin{equation}%
      \label{eq:model1Q1}%
      q_{_{\xi,\eta}}^{(\ell)} = \begin{cases} \sub{\alpha}{\ell} & \textrm{if  $\xi \prec \eta $} \\
        -\binom{|\xi|}{2}\sub{\alpha}{\ell} & \textrm{if $\xi = \eta$} \\
        0 & \textrm{otherwise}. \end{cases}%
      \end{equation}%
      in which $\xi \prec \eta $ now denotes a merger of at most two
      equivalence classes of $\xi$.  The usual Kingman coalescent is
      obtained from (\ref{eq:model1Q1}) by taking $X$ in
      (\ref{eq:defU}) to be a constant with value 1.  If large
      offspring number events are moderately frequent, ie.\ if
      $N^2\sub{\varepsilon}{N} \rightarrow \phi$ in which $\phi$ is a
      constant, the entries $ q_{_{\xi,\eta}}^{(\ell)}$ of the $Q_\ell$ matrix are 
      \begin{equation}%
         \label{eq:model1Q2}%
      q_{_{\xi,\eta}}^{(\ell)} = \begin{cases} \sub{\alpha}{\ell} +  \phi\mathbb{E}[Y^2(1 - Y)^{|\xi| - 2}]  & \textrm{if  $\xi \prec \eta $} \\
        \phi\mathbb{E}[Y^k(1 - Y)^{|\xi| - k}]  & \textrm{if $\xi \curlyeqprec \eta$, $ 3 \le k \le |\xi|$} \\
        -\sub{\alpha}{\ell}\binom{|\xi|}{2} - \phi\mathbb{E}[1 - (1 - Y)^{|\xi|} - |\xi|Y(1 - Y)^{|\xi| - 1}]  & \textrm{if $\xi = \eta$} \\
        0 & \textrm{otherwise}. \end{cases}%
        \end{equation}%
        in which $\xi \curlyeqprec \eta$ denotes a merger of at least
        three equivalence classes of $\xi$.  Finally, if
        $N^2\sub{\varepsilon}{N} \rightarrow \infty$ then large
        offspring number events occur quite frequently, and only one
        $Q$ matrix is obtained, whose entries are
        \begin{equation}%
         \label{eq:model1Q3}%
      q_{_{\xi,\eta}} = \begin{cases} \mathbb{E}[Y^k(1 - Y)^{|\xi| - k}]  & \textrm{if  $\xi \prec \eta $, $2 \le k \le |\xi|$} \\
        -\mathbb{E}[1 - (1 - Y)^{|\xi|} - |\xi|Y(1 - Y)^{|\xi| - 1}] & \textrm{if $\xi = \eta$ }\\
        0 & \textrm{otherwise}. \end{cases}%
        \end{equation}%

\subsection{Eigenvectors of the rate matrix associated with $A_t$ }%
\label{sec:eigenv}%
The eigenvectors of the rate matrix associated with either $A_t$ or
$\tilde{A}_t$ under the Kingman coalescent can be obtained in closed
form \citep{tavare84}.  The rate matrix associated with $A_t$ when
associated with a $\Lambda$ coalescent is triangular.  The
eigenvectors can be computed recursively as follows.  Let $\ell^{(k)}
= \left( \sub{\ell}{1}^{(k)}, \ldots, \sub{\ell}{n}^{(k)} \right)$ and $r^{(k)} =
\left( \sub{r}{1}^{(k)}, \ldots, \sub{r}{n}^{(k)} \right)$ denote the left and right
eigenvectors, respectively, corresponding to eigenvalue $\sub{\lambda}{k} =
\sub{q}{{k,k}}$.  Then $\sub{\ell}{i}^{(1)} = \delta_{1i}$, $\sub{\ell}{j}^{(k)} = 0$ if $j
> k$, $\sub{\ell}{k}^{(k)} = 1$ and, writing $q_{_k} \equiv q_{_{k,k}}$,
\begin{equation}%
  \label{eq:leigenv}%
  \sub{\ell}{j}^{(k)} =  \frac{ \sub{q}{{j+1,j}}\sub{\ell}{{j + 1}}^{(k)} + \cdots + \sub{q}{{k,j}}\sub{\ell}{k}^{(k)} }{\sub{q}{{k}} -  \sub{q}{{j}}}, \quad 1 \leq j < k.
  \end{equation}%
  Considering the right eigenvectors $r^{(k)}$, we have $r^{(1)} = (1,
  \ldots, 1)$, $r_k^{(k)} = 1$, $r_j^{(k)} = 0$ if $j < k$, and
\begin{equation}%
  \label{eq:reigenv}%
  \sub{r}{j}^{(k)} =  \frac{ \sub{q}{{j,k}}\sub{r}{k}^{(k)} + \cdots + \sub{q}{{j,j-1}}\sub{r}{{j-1}}^{(k)} }{ \sub{q}{{k}} - \sub{q}{{j}}},\quad  1 < k < j \leq n
  \end{equation}%
  One confirms that, for sample size two, $\ell^{(1)} =
  (1,0)$, $\ell^{(2)} = (-1,1)$, $r^{(1)} =
  (1,1)$, $r^{(2)} = (0,1)$, yielding $\mathbb{P}[A_t = 1
  | A_0 = 2] = 1 - \mathbb{P}[ A_t = 2 | A_0 = 2] = 1 -  e^{t\sub{q}{2}}$.

  The eigenvectors of the rate matrix associated with $\tilde{A}_t$
  when associated with a $\Lambda$ coalescent can similarly be
  obtained recursively (Equations~\ref{eq:leigenv}--\ref{eq:reigenv}),
  with $\ell^{(0)} = \delta_{0j}$, and $r^{(0)} = (1, \ldots, 1)$.

\subsection{The subsample coalescent process}

In this section we give the generators associated with the coancestry
coalescent process, when the ancestry of a subsample is considered
jointly with the ancestry of the sample. The convergence follows from
the convergence result obtained in the first section of the Appendix.
The generators will have similar form, though not quite the same, as
the one associated with the gene genelogical coancestry process
discussed in the main text.

Let $R_1$ and $R_2$ denote the sample and the subsample process,
respectively, on E.  To specify the joint process $\left(R_1(t),
  R_2(t); t \geq 0 \right)$, denote by $\xi \equiv (\xi_1, \xi_2)$ and
$\eta \equiv (\eta_1, \eta_2) \in E^2$ the states of the joint
process; with $\xi_i$ and $\eta_i$ denoting the states of $R_i$ for $i
\in \{1,2\}$.  By $\xi \preceq \eta$ we denote the transition
  \begin{displaymath}%
    \xi \preceq \eta \equiv \begin{cases}  \xi_1 \prec \eta_1, \quad \xi_2 \prec \eta_2 \\
      \xi_1 \prec \eta_1, \quad \xi_2 = \eta_2. \\
      \end{cases}
    \end{displaymath}%
    where $\xi_1 \prec \eta_1$ denotes a merger of two
    classes of $\xi_1$.   By $\xi \curlyeqprec \eta$ we denote
    the transition
    \begin{displaymath}%
      \xi \curlyeqprec \eta \equiv \begin{cases}   \xi_1 \curlyeqprec \eta_1, \quad \xi_2 \curlyeqprec \eta_2 \\
         \xi_1 \curlyeqprec \eta_1, \quad \xi_2 \prec \eta_2 \\
          \xi_1 \curlyeqprec \eta_1, \quad \xi_2 = \eta_2 \\
        \end{cases}%
      \end{displaymath}%
      in which $\xi_1 \curlyeqprec \eta_1$ denotes a merger of three
      or more classes of $\eta_1$, which then can involve more than
      two classes $(\xi_2 \curlyeqprec \eta_2)$, exactly two $(\xi_2
      \prec \eta_2)$, or at most one $(\xi_2 = \eta_2)$ class of the
      subsample process.

      In discrete time, the joint process $\left( R_1(r),
        R_2(m) \right)_{m \in \mathbb{N}}$ is a Markov chain with
      transition probabilities $P_{\xi,\eta}(N)$ determined by
  \begin{equation}%
  P_{\xi, \eta}(N)  = \frac{\mathbb{E}[(U)^{k}(N - U - 1)_{i-k}]}{(N)_i} \frac{(|\xi_1| - |\xi_2|)_{k - \ell}(|\xi_2|)_\ell }{(|\xi_1|)_k }
    \end{equation}%
    in which $U^{(k)} \equiv (U)_k + (U)_{k - 1}$.  The results of
    previous sections now give convergence of finite-dimensional
    distributions of $\left(R_1(\lfloor t\sub{T}{N} \rfloor ), R_2
      (\lfloor t \sub{T}{N} \rfloor) \right)$ to $\left(R_1(t), R_2(t)
    \right)$.  If $N^2\sub{\varepsilon}{N} \rightarrow 0$ large
    offspring number events are negligible and the rate matrix
    $Q_\ell$ of $\left(R_1(t), R_2(t) \right)$ associated with model
    $\ell \in \{1,2\}$ has entries
    \begin{equation}%
       \sub{q}{{\xi, \eta}}^{(\ell)} = \begin{cases}  \sub{\alpha}{\ell}\frac{(|\xi_1| - |\xi_2|)_{2 - j}(|\xi_2|)_j }{(|\xi_1|)_2 }   &  \xi \preceq \eta, \quad 0 \le j  \le 2\wedge  |\xi_2|  \\\\ 
         -\sub{\alpha}{\ell}\binom{|\xi_1|}{2} & \xi = \eta. \\
         \end{cases}%
      \end{equation}%
      in which $\sub{\alpha}{\ell}$ are given in Equation~(\ref{eq:alphadef}). 
      If $N^2\sub{\varepsilon}{N} \rightarrow \phi$ where $\phi > 0$
      is a constant, the rate matrix associated with model $\ell$  has entries
       \begin{equation}%
       \sub{q}{{\xi, \eta}}^{(\ell)} = \begin{cases}  \frac{(|\xi_1| - |\xi_2|)_{2 - j}(|\xi_2|)_j }{(|\xi_1|)_2 }\left(\sub{\alpha}{i}  +  \phi\mathbb{E}[Y^2(1 - Y)^{|\xi_1| - 2}]  \right)   &  \xi \preceq \eta, \quad 0 \le j  \le 2\wedge |\xi_2|  \\\\ 
         \phi\mathbb{E}[Y^k(1 - Y)^{|\xi_1| - k}] \frac{(|\xi_1| - |\xi_2|)_{k - j}(|\xi_2|)_j }{(|\xi_1|)_k } &  3 \le k \le |\xi_1|,\quad 0 \le j \le  k\wedge |\xi_2|   \\\\
    -\sub{\alpha}{\ell}\binom{|\xi_1|}{2} - \phi\mathbb{E}[1 - (1 - Y)^{|\xi_1|} - |\xi_1|Y(1 - Y)^{|\xi_1| - 1}]    & \xi = \eta. \\
         \end{cases}%
      \end{equation}%
      If $N^2\sub{\varepsilon}{N} \rightarrow 0$ large offspring
      number events are dominating and only one rate matrix is
      possible, whose entries are
     \begin{equation}%
       \sub{q}{{\xi, \eta}} = \begin{cases} 
         \mathbb{E}[Y^k(1 - Y)^{|\xi_1| - k}] \frac{(|\xi_1| - |\xi_2|)_{k - j}(|\xi_2|)_j }{(|\xi_1|)_k } &  \xi  \curlyeqprec  \eta ,\quad 2 \le k \le |\xi_1|,\\
&  \quad 0 \le j \le  k\wedge |\xi_2|;   \\\\
     -\mathbb{E}[1 - (1 - Y)^{|\xi_1|} - |\xi_1|Y(1 - Y)^{|\xi_1| - 1}]    & \xi = \eta. \\
         \end{cases}%
      \end{equation}%

\subsection{The conditional distributions of $A_1$ and $A_2$ }
\label{sec:subsample}

In this section we obtain the conditional distribution of the number
of ancestors $A_2$ of the subsample, given the number of ancestors
$A_1$ of the sample.  The conditional distribution of $A_2$, and the
results of section (\ref{sec:eigenv}) then yield the joint
distribution of $A_1$ and $A_2$.  The results for the subsample
process are straightforward extensions of those obtained for the
Kingman coalescent \citep{saunders84}.  To keep notation in line with
\cite{saunders84}, let $A_1(t)$ denote the Markov chain
counting the number of distinct ancestral lineages at time $t \ge 0$
before the present of a set of lineages drawn from the population at
time zero.  Let $A_2(t)$ denote the corresponding number for a subset
of the lineages drawn at time zero.  Define $T_1(a) \equiv
\inf\left\{t: A_1(t) = a\right\}$, $T_2(a) \equiv
\inf\left\{t: A_2(t) = a\right\}$, $A_1^*(a)
\equiv A_1(T_2(a))$, and $A_2^{*}(a) \equiv A_2(T_1(a))$.  Given $A_1(t) = a$, denote by
$\beta(a, a - k)$ the probability that the next merger involves $k +
1$ active ancestral lineages of the sample.  The process $\left(
  A_2^*(a - \ell_1), \ell_1 = 0, \ldots, a - 1 \right)$ is a Markov
chain with transition probabilities
\begin{equation}%
  \label{eq:PA2star}%
  \mathbb{P}\left[ A_2^*(\sub{a}{1} - k) = \sub{a}{2} - j | A_2^*(\sub{a}{1}) = \sub{a}{2} \right]  = \frac{\binom{\sub{a}{2}}{j + 1}\binom{\sub{a}{1} - \sub{a}{2}}{k - j} }{\binom{ \sub{a}{1}}{k + 1} } 
  \end{equation}%
  for $ 1
  \leq j \leq \min(k, \sub{a}{2} - 1)$, and $ 1 \leq k \leq \sub{a}{1} - 1$, 
and marginal probabilities  
\begin{equation}%
\mathbb{P}\left[ A_2^*(\sub{a}{1}) = \sub{a}{2} | A_2^*(\sub{a}{1}) = \sub{a}{2} \right] = 1, \quad 1 \leq \sub{a}{2} \leq \sub{a}{1}.   
  \end{equation}%
  The transition probability (\ref{eq:PA2star}) is not a function of
  the coalescence rates, since we condition on a merger of $k + 1$
  lineages.  However, the conditional
  distribution $$\phi_{ij}(\sub{a}{1},\sub{a}{2}) \equiv
  \mathbb{P}\left[ A_2^*(\sub{a}{1}) = \sub{a}{2} | A_2(0) = j, A_1(0)
    = i \right]$$ of $A_2$, given $A_1$, will be, due to the presence
  of multiple mergers. Write $\phi(\cdot,\cdot) =
  \phi_{i,j}(\cdot,\cdot)$ for ease of presentation.  The conditional
  distribution of $A_2^*$, given $A_1$, can be obtained as a recursion,
\begin{equation}%
  \label{eq:phiA2A1}
  \begin{split}
  \phi(\sub{a}{1}, \sub{a}{2})&  = \sum_{k = 1}^{i - a}\beta(\sub{a}{1} + k, \sub{a}{1})\sum_{\ell = 1}^{k\wedge j}\left(  \phi(\sub{a}{1} + k, \sub{a}{2} + \ell)\frac{   \binom{\sub{a}{2} + \ell }{\ell + 1 }\binom{\sub{a}{1} + k - \sub{a}{2} - \ell }{k - \ell }  }{\binom{\sub{a}{1} + k }{k +1 } }   \right. \\\\
& \quad +  \left.   \phi( \sub{a}{1} + k, \sub{a}{2} ) \frac{ \sub{a}{2} \binom{\sub{a}{1} + k - \sub{a}{2} }{k} + \binom{\sub{a}{1} + k - \sub{a}{2} }{k + 1} }{\binom{\sub{a}{1} + k }{k + 1 } }    \right)  \\
  \end{split}%
  \end{equation}%
  with the boundary conditions $\phi_{i,j}(i,j) = 1$ and
  $\phi_{i,1}(i,1) = 1$ for all $i \geq 1$.  By way of example,
  $\phi_{1,1}(1,1) = \phi_{2,1}(2,1) = \phi_{2,1}(1,1) = 1$ and
  $\phi_{3,2}(2,1) = \beta(3,2)/3$.  The joint distribution of
  $(A_1(t), A_2(t))$ is 
\begin{equation}%
  \mathbb{P}\left[ A_1(t) = a_1,  A_2(t) = a_2 | A_1(0) = i, A_2(0) = j \right] =  g_{i,{\sub{a}{1}}}(t)\phi_{i,j}(\sub{a}{1}, \sub{a}{2})
  \end{equation}%
  in which $\phi_{i,j}(\sub{a}{1}, \sub{a}{2})$ is given in
  Equation~(\ref{eq:phiA2A1}).  
 
  The distribution of $A_1^*(\cdot)$ can be given via a simple
  recursion,
\begin{multline}%
  \mathbb{P}\left[ A_1^*(\sub{a}{2}) = \sub{a}{1} | A_1(0) = i, A_2(0) = j \right] = \\ \sum_{k = 1}^{i - a_1} \beta( \sub{a}{1} + k, \sub{a}{1}) \sum_{\ell=1}^{k\wedge (j - a_2)}     \phi_{i,j}(\sub{a}{1} + k, \sub{a}{2} + \ell)\frac{ \binom{ \sub{a}{2} + \ell }{\ell + 1 }\binom{ \sub{a}{1} + k - \sub{a}{2} - \ell }{k - \ell }}{\binom{ \sub{a}{1} + k }{k + 1 } }.
\end{multline} %

\subsection{The subsample coalescent process including mutation}

In this section we give the generators associated with the coancestry
coalescent process including neutral mutation.  Let $\tilde{R}_1$ and
$\tilde{R}_2$ denote the coalescent processes for the sample and the
subsample, respectively, including mutation.  As we are assuming an
infinite allele/sites mutation model, mutation results in `new'
alleles, and the probability of recurrent mutation is ignored.  The
joint process takes values in $\tilde{E}^2$, in which the transition
$\xi \rightsquigarrow \eta$, as previously, denotes a transition due
to mutation;
  \begin{equation}%
    \label{eq:jointmuttransi}%
    \xi \rightsquigarrow \eta = \begin{cases}  \xi_1 \rightsquigarrow \eta_1, & \xi_2 = \eta_2; \\
      \xi_1 \rightsquigarrow \eta_1, & \xi_2 \rightsquigarrow \eta_2. \\
      \end{cases}
    \end{equation}%
    The first line in (\ref{eq:jointmuttransi}) denotes the transition
    in which the mutated lineage does not belong to the subsample,
    while the second line means that the mutated lineage does belong
    to the subsample.  The set $\tilde{E}$ denotes the space of all
    equivalence relations on $\{1, \ldots, n\}$ in which each relation $\xi$ is
    partitioned into a set of new and old equivalence classes \citep{mohle99weak}.   
 
    The joint ancestral process $\left(\tilde{R}(k); k \in
      \mathbb{N}\right) \equiv \left(\tilde{R}_1(k),
      \tilde{R}_2(k)\right)_{k \in \mathbb{N}}$ is a Markov chain on
    $\tilde{E}^2$ with transition probability
    $\tilde{P}_{\xi,\eta}(N)$ given by, in which $m$ denotes the
    number of classes that mutate and $m + b$ is the number of old classes of
    $\xi$,
  \begin{equation}%
    P_{\xi,\eta}(N) = \mu_{_N}^m(1 - \mu_{_N})^b\frac{\mathbb{E}[U^{(k)}(N - 1 - U)_{k - b}]}{(N)_b} \frac{(|\xi_1| - |\xi_2|)_{k - j}(|\xi_2|)_j }{(|\xi_1|)_k },
    \end{equation}%
    in which $2 \le k \le |\xi_1|$, and $0 \le j \le k\wedge
    |\xi_2|$.    The joint process $\tilde{R}\left(\lfloor tT_N \rfloor
    \right)$ associated with model $\ell \in \{1,2\}$ converges weakly to $\tilde{R}(t)$ with generator
    $\tilde{Q}_\ell \equiv \left(\tilde{q}_{_{\xi,\eta}}^{(\ell)} \right)_{\xi,\eta
      \in \tilde{E}^2}$.  If $N^2\sub{\varepsilon}{N} \rightarrow 0$
    large offspring number events are negligible and the generator
    associated with model $\ell \in \{1,2\}$ has entries
    \begin{equation}%
       \tilde{q}_{_{\xi,\eta}}^{(\ell)} = \begin{cases} \theta/2 & \xi \rightsquigarrow \eta \\\\
          \sub{\alpha}{\ell}\frac{(|\xi_1| - |\xi_2|)_{2 - j}(|\xi_2|)_j }{(|\xi_1|)_2 }  &  \xi \preceq \eta, \quad  0 \le j \le 2\wedge |\xi_2| \\\\
      -|\xi_1|\frac{\theta}{2}  - \binom{|\xi_1|}{2}\sub{\alpha}{\ell}   & \xi = \eta. \\
         \end{cases}
      \end{equation}%
      in which the constants $\sub{\alpha}{\ell}$ are given in Equation~(\ref{eq:alphadef}).
      If $N^2\sub{\varepsilon}{N} \rightarrow \phi$ where $\phi > 0$
      is a constant, the generator associated with model $\ell$ has entries  
      \begin{equation}%
       \tilde{q}_{_{\xi,\eta}}^{(\ell)} = \begin{cases} \theta/2 & \xi \rightsquigarrow \eta \\\\
         \left(\sub{\alpha}{\ell} +  \phi\mathbb{E}[Y^2(1 - Y)^{|\xi_1| - 2}]\right) \frac{(|\xi_1| - |\xi_2|)_{2 - j}(|\xi_2|)_j }{(|\xi_1|)_2 }  &  \xi \preceq \eta, \quad  0 \le j \le 2\wedge |\xi_2| \\\\
          \phi\mathbb{E}[Y^k(1 - Y)^{|\xi_1| - k}]\frac{(|\xi_1| - |\xi_2|)_{k - j}(|\xi_2|)_j }{(|\xi_1|)_k }   &    \xi \curlyeqprec \eta, \quad 3 \le k \le |\xi_1| \\\\
      -|\xi_1|\frac{\theta}{2}  - \binom{|\xi_1|}{2}\sub{\alpha}{\ell} - \phi\mathbb{E}[1 - (1 - Y)^{|\xi_1|} - |\xi_1|Y(1 - Y)^{|\xi_1| - 1}]  & \xi = \eta. \\
         \end{cases}
      \end{equation}%
      If $N^2\sub{\varepsilon}{N} \rightarrow \infty$ large offspring
      number events are dominating and the single resulting generator has entries
      \begin{equation}%
         \tilde{q}_{_{\xi,\eta}} = \begin{cases} \theta/2 & \xi \rightsquigarrow \eta \\\\
            \mathbb{E}[Y^2(1 - Y)^{|\xi_1| - 2}] \frac{(|\xi_1| - |\xi_2|)_{2 - j}(|\xi_2|)_j }{(|\xi_1|)_2 }  &  \xi \preceq \eta, \quad  0 \le j \le 2\wedge |\xi_2| \\\\
              \mathbb{E}[Y^k(1 - Y)^{|\xi_1| - k}]\frac{(|\xi_1| - |\xi_2|)_{k - j}(|\xi_2|)_j }{(|\xi_1|)_k }   &    \xi \curlyeqprec \eta, \quad 3 \le k \le |\xi_1| \\\\
               -|\xi_1| \frac{\theta}{2} - \mathbb{E}[1 - (1 - Y)^{|\xi_1|} - |\xi_1|Y(1 - Y)^{|\xi_1| - 1}]  & \xi = \eta. \\
               \end{cases}%
        \end{equation}%

\subsection{The conditional distributions of $\tilde{A}_1^*$ and  $\tilde{A}_2^*$  }

      The conditional distribution of $\tilde{A}_2^*(\cdot)$, given
      initial subsample size $j$ from initial sample size $i$, is
      obtained similarly as the conditional distribution of
      $A_2^*(\cdot)$.  Define
\begin{displaymath}%
  \tilde{\phi}_{i,j}(a_1, a_2) \equiv  \mathbb{P}\left[ \tilde{A}_2^*(a_1) = a_2 | \tilde{A}_2(0) = j , \tilde{A}_1(0) = i \right],
\end{displaymath}%
let $\lambda(a)$ denote the transition rate of the sample process
$\tilde{A}_1$ given $a$ active ancestral lineages, and let $\beta(a +
k, a)$ denote the probability of coalescence of $k+1$ out of $a + k$
active ancestral lineages.  The forward equations of
$\tilde{\phi}(\cdot,\cdot) = \tilde{\phi}_{i,j}(\cdot,\cdot)$ are
given by
\begin{equation}%
\begin{split}%
  \tilde{\phi}(a_1, a_2)& =  \tilde{\phi}(a_1 + 1, a_2)\left(\tilde{\beta}(a_1 + 1, a_1)\frac{(a_1 - a_2 + 1)(a_1 + a_2)}{(a_1 + 1)a_1 }  +  \frac{(a_1 + 1)\theta/2}{\lambda(a_1 + 1)}\frac{a_1 - a_2 + 1 }{1 + a_1} \right)  \\\\
& \quad + \tilde{\phi}(a_1 + 1, a_2 + 1)\left(\tilde{\beta}(a_1 + 1, a_1)\frac{(a_2 + 1)a_2}{(a_1 + 1)a_1} + \frac{(a_1 + 1)\theta/2 }{\lambda(a_1 + 1)}\frac{a_2 + 1}{a_1 + 1} \right) \\\\
& \quad + \sum_{k = 2}^{i - a_1} \tilde{\phi}(a_1 + k, a_2)\tilde{\beta}(a_1 + k, a_1)\frac{\binom{a_1 + k - a_2 }{k + 1} }{\binom{a_1 + k }{k + 1} } \\\\
& \quad + \sum_{k = 2}^{i - a_1} \tilde{\phi}(a_1 + k, a_2)\tilde{\beta}(a_1 + k, a_1)\frac{ a_2\binom{ a_1 + k - a_2  }{k }  }{ \binom{a_1 + k }{k + 1}  } \\\\
& \quad  + \sum_{k = 2}^{i - a_1}\sum_{\ell = 2}^{k\wedge(j - a_2)}\tilde{\phi}(a_1 + k, a_2 + \ell)\tilde{\beta}(a_1 + k, a_1)\frac{\binom{a_2 + \ell - 1 }{\ell }\binom{a_1 + k - a_2 - \ell }{k - \ell }  }{\binom{a_1 + k}{k + 1 }  }   \\\\
  \end{split}%
  \end{equation}%
  with boundary condition $\tilde{\phi}_{i,k}(i,k) = \delta_{k,j}$. 

The conditional distribution of $\tilde{A}_1^*$ can be obtained similarly.  Define, 
\begin{equation} 
\tilde{\psi}_{i,j}(a_1, a_2) \equiv \mathbb{P}\left[\tilde{A}_1^*(a_2) = a_1 | \tilde{A}_1(0) = i, \tilde{A}_2(0) = j \right].  
\end{equation}
The forward equations for $\tilde{\psi}(\cdot, \cdot) =
  \tilde{\psi}_{i,j}(\cdot,\cdot)$ are given by  
\begin{equation}%
  \begin{split}%
    \tilde{\psi}(a_1, a_2) & = \tilde{\psi}(a_1 + 1, a_2 + 1)\left( \beta(a_1 + 1, a_1) \frac{(a_2 + 1)a_2}{(a_1 + 1)a_1} +  \frac{(a_1 + 1)\theta/2}{\lambda(a_1 + 1)}\frac{a_2 + 1}{a_1 + 1} \right) \\\\ 
    & + \sum_{k = 2}^{i - a_1} \beta(a_1 + k, a_1) \sum_{\ell = 2}^{k \wedge (j - a_2)}\tilde{\psi}(a_1 + k, a_2 + \ell) \frac{ \binom{a_2 + \ell }{\ell + 1 }\binom{a_1 + k  - a_2 - \ell }{k - \ell }  }{\binom{a_1 + k}{k + 1}} 
    \end{split}%
  \end{equation}%
  with boundary conditions $\tilde{\psi}_{i,j}(i,j) =
  \tilde{\psi}_{i,i}(\cdot,\cdot) = 1$.   By way of example,
\begin{equation}%
  \tilde{\psi}_{2,1}(0,0) =  \beta(2,1) + \frac{\theta }{4\lambda(2)}
\end{equation}

\clearpage
\pagebreak
\newpage

\begin{table}[H!]
  \caption{An approximation of the probability  $h_{j}(t)$ (\ref{eq:hjt}) that $j$ of the original lines in the population are still segregating at time  $t = 1$ for  different values of $\theta$ and  $\beta_\ell$.  Results for the usual Kingman coalescent are obtained when $\beta_\ell = 1$.    }%
 \label{tab:tab1}%
 \centering
 \begin{tabular}{llllll}%
   \hline
   & & \multicolumn{3}{c}{j} \\
     $\theta$ &  $\beta_\ell$ & 0 & 1 & $ \ge 2$ \\
   \hline 
      1 & 1 & $0.036$ & $0.320$ & $0.644$ \\
        & 2 & $0.133$ & $0.654$ & $0.213$ \\
        & 5 & $0.273$ & $0.718$ & $0.009$ \\
      5 & 1 & $0.556$ & $0.379$ & $0.065$ \\
        & 2 & $0.719$ & $0.270$ & $0.011$ \\
        & 5 & $0.836$ & $0.164$ & $0.000$ \\
   \hline
   \end{tabular}%
   \end{table}%

\pagebreak
\clearpage
\newpage

\begin{table}[H!]
  \caption{ The probability $\psi_{i,2}(1,1)$ (Equation~\ref{eq:psiij11}) of two lineages sharing a most recent common ancestor with the $i$ lineages, varying over $i$ and $\pi$.  When $0 < \pi < 1$, $\pi$ denotes  $\psi$ of the point mass process;  $\pi$ denotes $\alpha$ of the beta coalescent when $1 < \pi < 2$,   and finally the vector $(\alpha,\beta)$ of the two-parameter beta coalescent.  }%
\label{tab:table2}%
\centering
\begin{tabular}{rcr}%
  \hline
  $i$ &  $\pi$ &  $\psi_{i,2}(1,1)$ \\
  \hline
  500 &  $0.005$ & $0.330$ \\
      & $0.5$    & $0.065$ \\
  1000& $0.005$ & $0.328$ \\
      & $0.5$ & $0.048$ \\
  500 & $1.05$& $0.178$ \\
      & $1.5$ & $0.266$ \\
 1000 & $1.05$ & $0.163$ \\
      & $1.5$ & $0.261$ \\
 500  & $(1.0,1.0)$ & $0.061$ \\
      & $(1.0,5.0)$ & $0.122$ \\
 1000 & $(1.0,1.0)$ & $0.041$ \\
      & $(1.0,5.0)$ & $0.105$ \\
  \hline
  \end{tabular}\end{table}%
\pagebreak
\clearpage
\newpage

\begin{table}[H!]
  \caption{ The probability $\tilde{u}(1000,2)$ that the oldest allele of a sample is among two lineages drawn from a sample of size $1000$.  Cases of $\pi < 1$ refer to the point mass process,  of $1 < \pi < 2$ to the beta coalescent, and  $\pi = (\alpha, \beta)$ refers to the two-parameter beta coalescent.    }%
\label{tab:table3}%
\centering
\begin{tabular}{rcr}%
  \hline
  $\theta$ & $\pi$ & $\tilde{u}(1000,2)$ \\
 \hline
 1 & $1.05$ & $0.506$ \\
   & $1.5$ & $0.598$ \\
 5  & $1.05$ & $0.111$ \\
  &   $1.5$ & $0.189$ \\
 1 & $0.05$ & $0.003$ \\
   & $0.5$ & $0.071$ \\
 5 & $0.05$ & $0.002$ \\
   & $0.5$ & $0.008$ \\
 1 & $(1,1)$ & $0.148$ \\
   & $(1,5)$ & $0.010$ \\
 5 & $(1,1)$ & $0.022$ \\
   & $(1,5)$ & $0.003$ \\
 1 & $(5,1)$ & $0.354$ \\
 5 & $(5,1)$ & $0.071$ \\
 1 & $(5,5)$ & $0.084$ \\
 5 & $(5,5)$ & $0.010$ \\
  \hline
  \end{tabular}\end{table}%

\pagebreak
\clearpage
\newpage

\begin{table}[H!]
  \caption{ The mean $\mathbb{E}[F_i]$ and variance  $\mathbb{V}[F_i]$ of the number $F_i$ of the oldest allele  in a sample of size $i = 30$ varying over $\theta$ and $\pi$.  The corresponding values for the Kingman coalescent are  $\mathbb{E}[F_i] = 15.500$ and  $\mathbb{V}[F_i] = 74.917$ for $\theta = 1$, and   $\mathbb{E}[F_i] = 5.833$ and  $\mathbb{V}[F_i] = 20.139$ for $\theta = 5$.    }%
\label{tab:table3}%
\centering
\begin{tabular}{rcrr}%
  \hline
  $\theta$ & $\pi$ & $\mathbb{E}[F_i]$ & $\mathbb{V}[F_i]$ \\
 \hline
 1 & $1.05$ & $14.834$ &  $93.625$ \\
   & $1.5$  & $15.026$ &  $84.975$ \\
 5 & $1.05$ & $5.112$  &  $34.612$ \\
   & $1.5$ & $5.269$  & $26.698$  \\
 1 & $0.05$& $1.061$ &  $0.096$ \\
  & $0.5$ & $5.182$ & $38.028$ \\
 5 & $0.05$& $1.012$&  $0.018$ \\
  & $0.5$ & $1.824$&  $5.995$ \\
 1& $(1,1)$ & $7.792$ & $67.565$ \\
 & $(5,1)$ & $13.570$ & $96.908$ \\
 & $(1,5)$ & $1.845$&  $5.027$ \\
 & $(5,5)$& $5.752$& $45.621$ \\
5& $(1,1)$& $2.630$& $18.001$ \\
& $(5,1)$& $4.923$& $44.402$ \\
& $(1,5)$& $1.163$& $0.826$ \\
& $(5,5)$& $1.978$& $8.200$ \\ 
  \hline 
  \end{tabular}\end{table}%

\pagebreak
\clearpage
\newpage

Figure 1: The probability $h_{i0}(t)$ that $i$ lineages have vanished from the
population by time $t$ as a function of time for $i = 20$ and varying
over $\theta$ and the coalescence parameters $\psi$ (a), $\alpha$ (b),
and $(\alpha,\beta)$ (c) as shown in the legends. 

\begin{picture}(50,50)
\put(100,-135){\includegraphics[height=2.5in,width=2.5in,scale=1,angle=0]{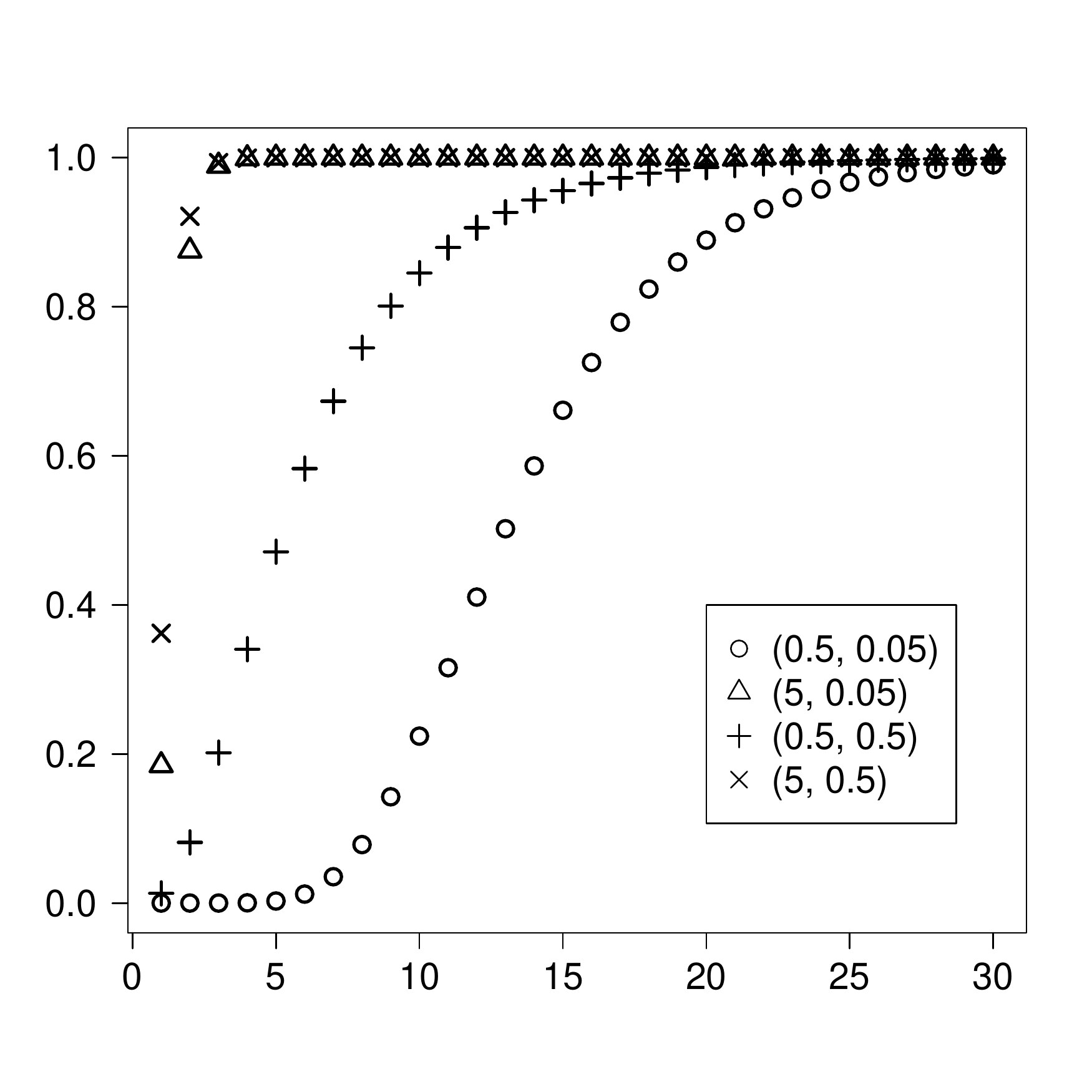}}%
\put(100,-320){\includegraphics[height=2.5in,width=2.5in,scale=1,angle=0]{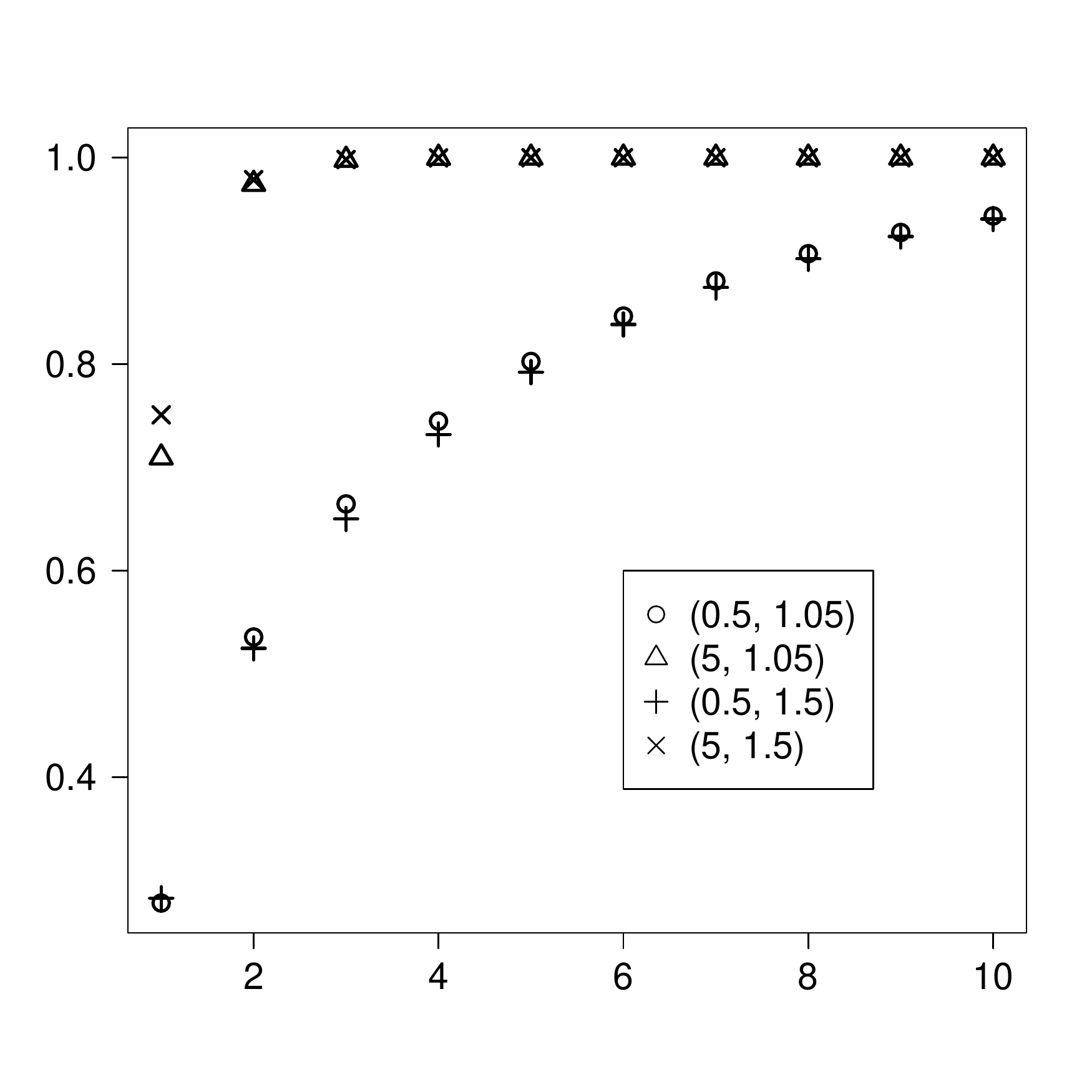}}%
\put(100,-500){\includegraphics[height=2.5in,width=2.5in,scale=1,angle=0]{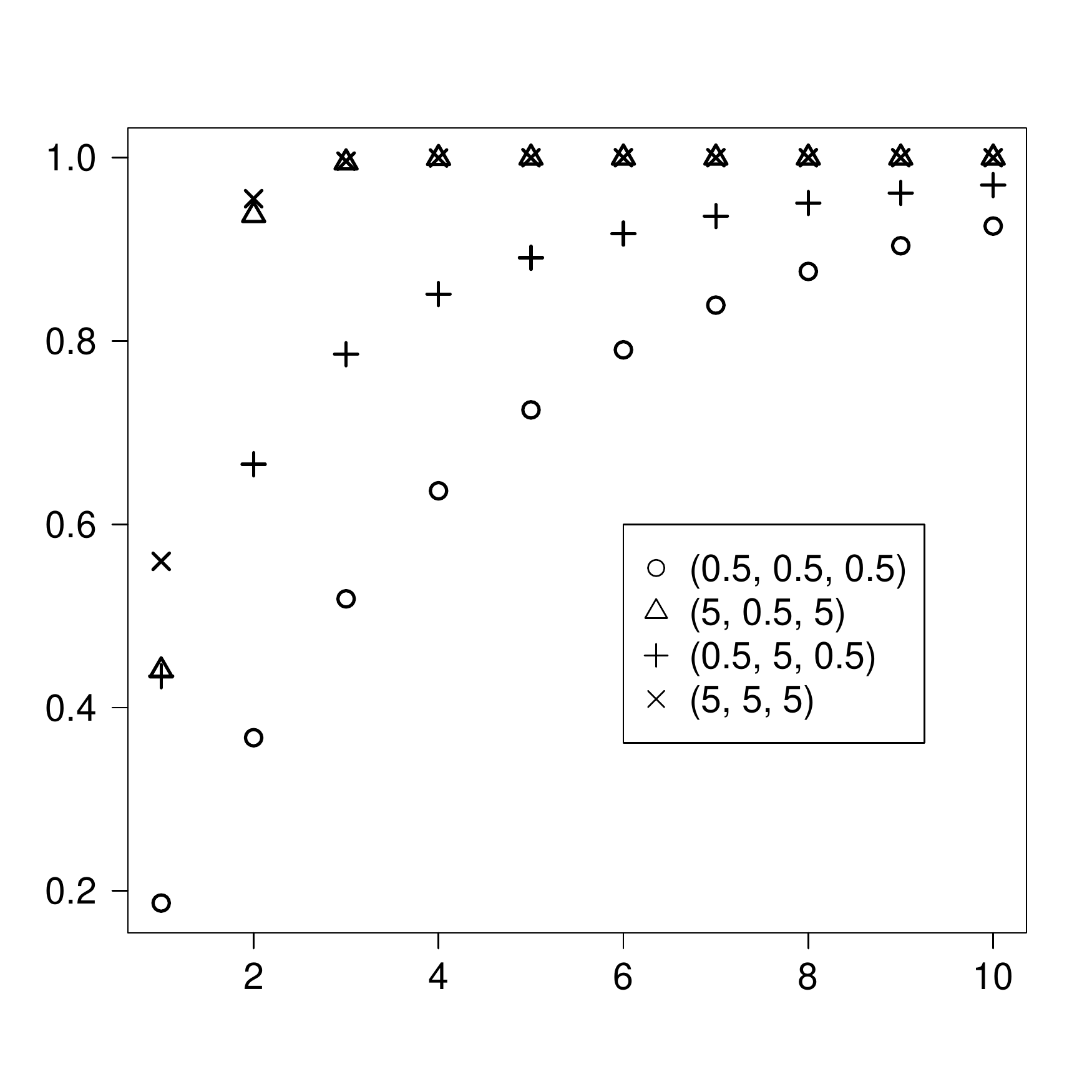}}%
\put(150,30){(a) $\psi$-coalescent}\put(150,-155){(b) beta coalescent}
\put(150,-335){(c) beta$(\alpha, \beta)$ coalescent }%
\put(224,-50){$(\theta,\psi)$}\put(220,-230){$(\theta,\alpha)$}%
\put(217,-400){$(\theta,\alpha,\beta)$}
\put(170,-496){time $t$}%
\end{picture}%

\pagebreak
\clearpage
\newpage


Figure 2: Expected number $\mathbb{E}[\sub{\tilde{A}^*}{2}(\sub{a}{1}) =
\sub{a}{2} | \sub{\tilde{A}}{1} = 150, \sub{\tilde{A}}{2} = 50]$ of
lineages ancestral to the subsample as a function of $\sub{a}{1}$ when
the population has the $\alpha$-coalescent with $\alpha$ and the
mutation rate $\theta$ varying as shown in the legend.  Right panels
are for the two parameter beta coalescent.

\begin{picture}(50,50)
  \put(0,-140){\includegraphics[height=2.5in,width=2.5in,scale=1,angle=0]{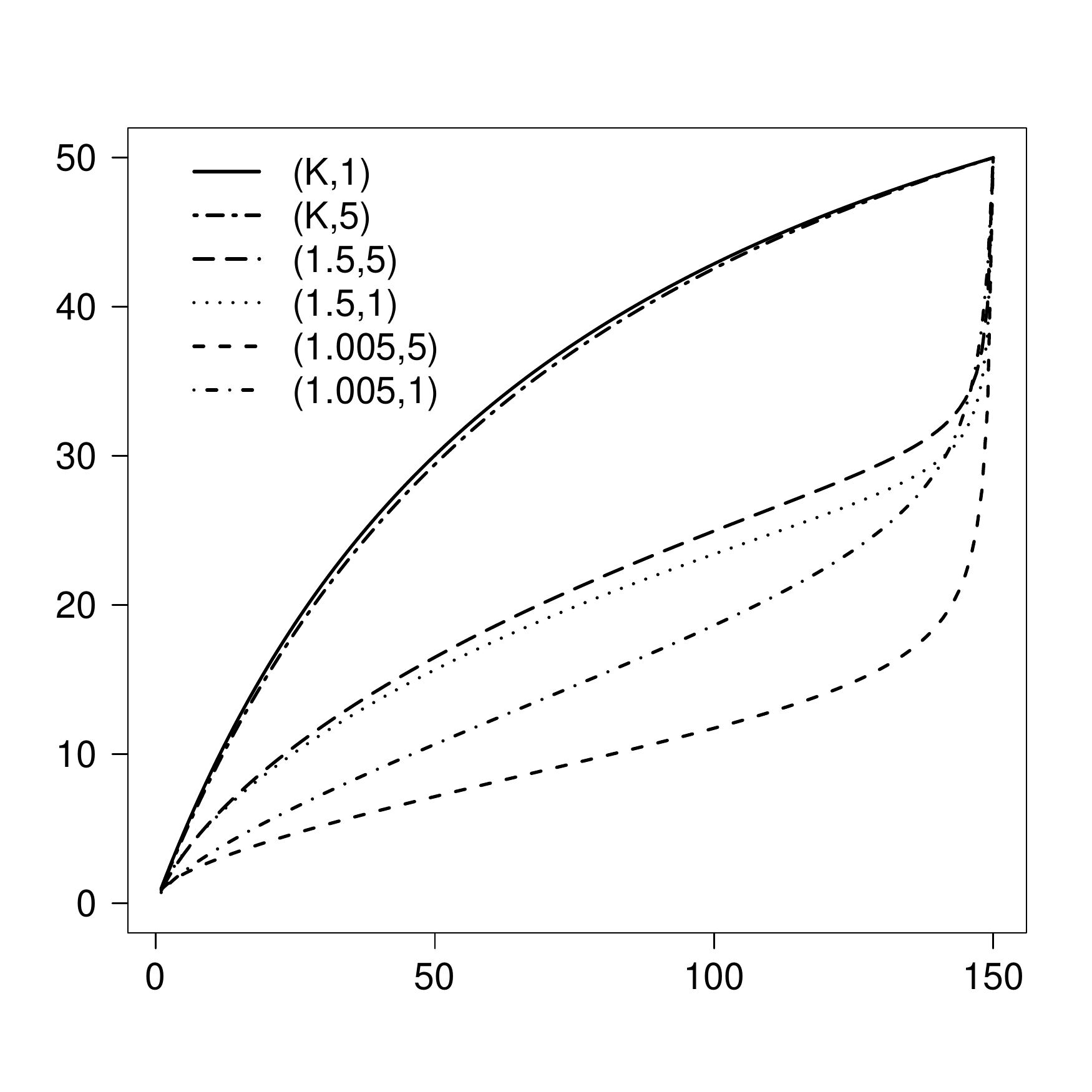}}%
  \put(0,-330){\includegraphics[height=2.5in,width=2.5in,scale=1,angle=0]{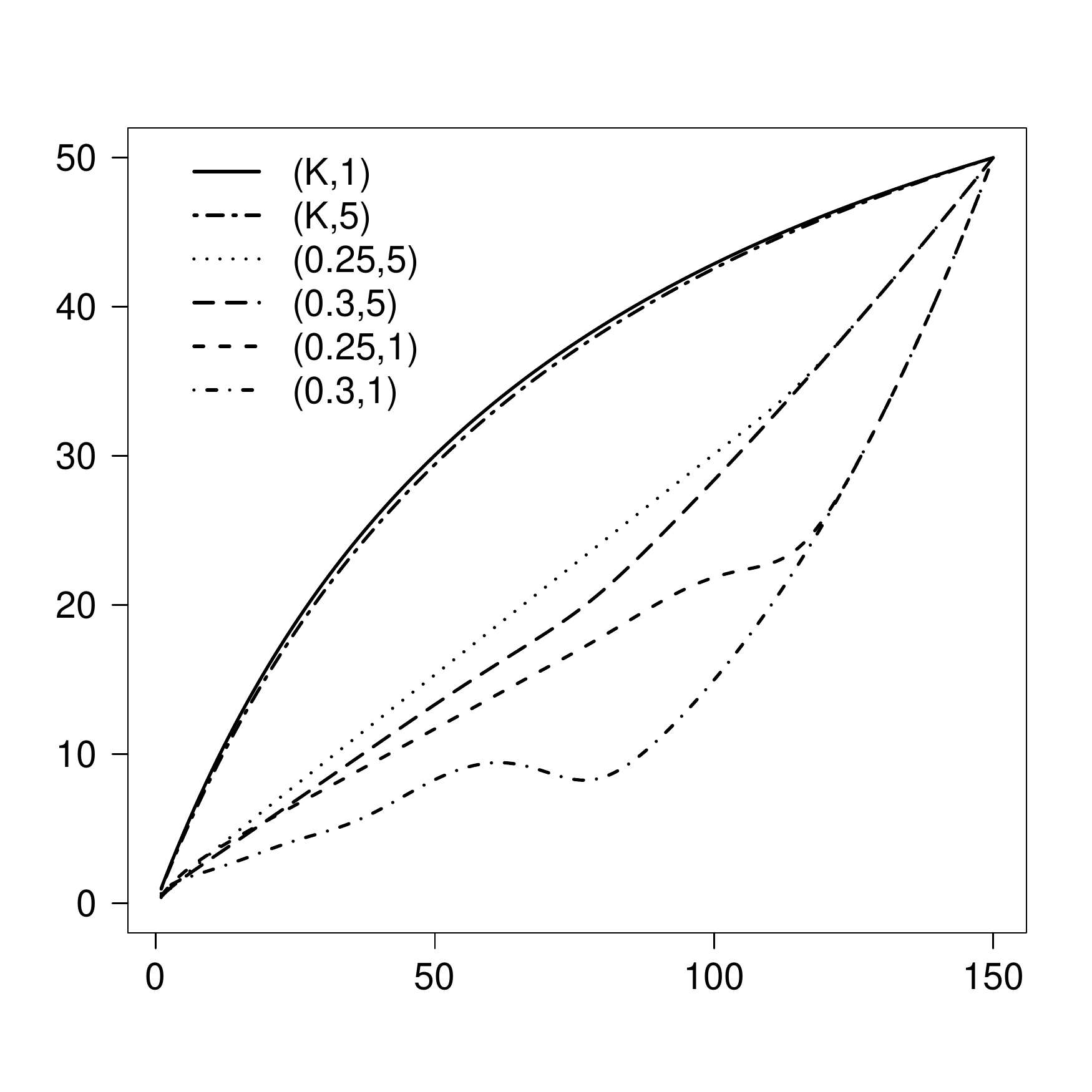}}%
  \put(200,-140){\includegraphics[height=2.5in,width=2.5in,scale=1,angle=0]{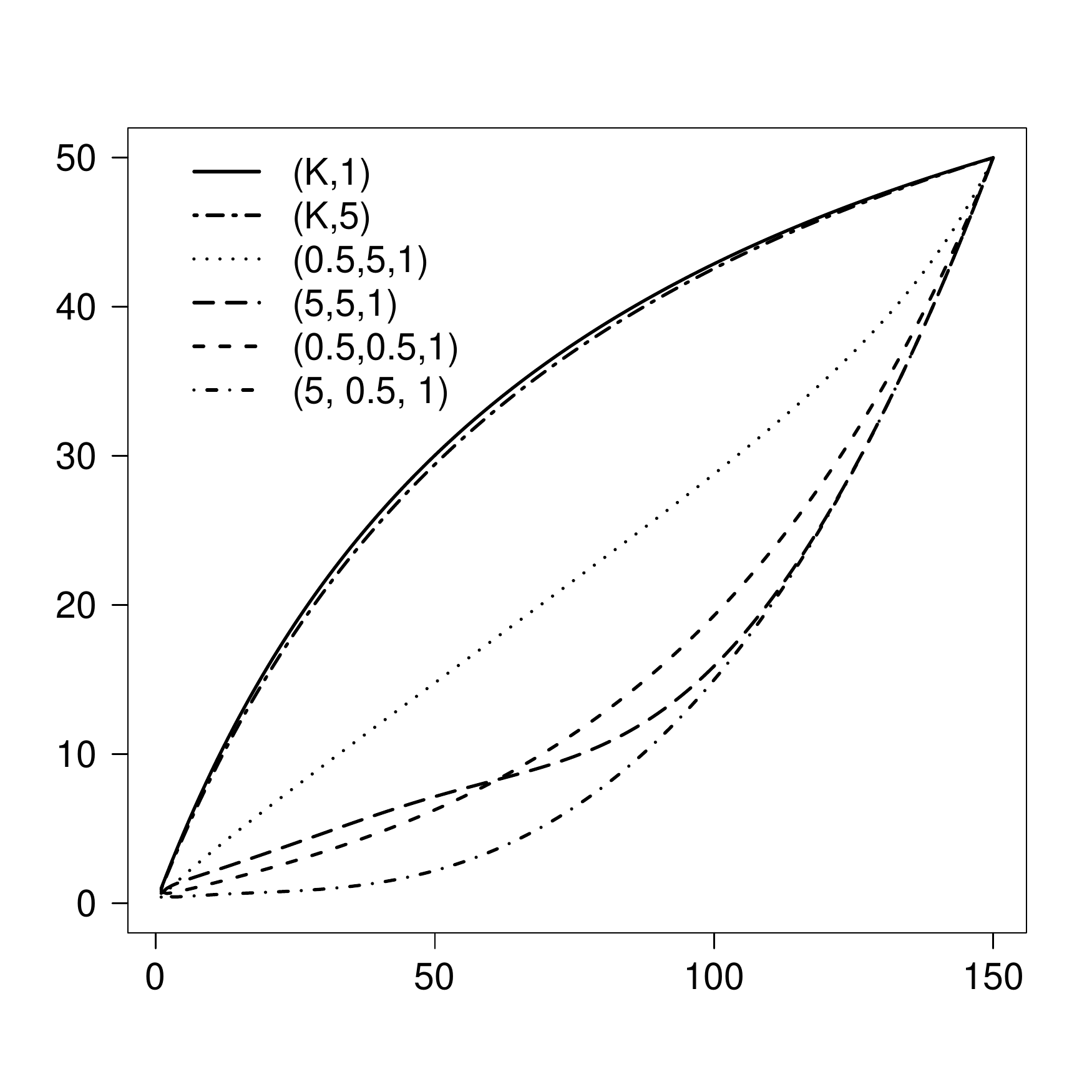}}%
  \put(200,-330){\includegraphics[height=2.5in,width=2.5in,scale=1,angle=0]{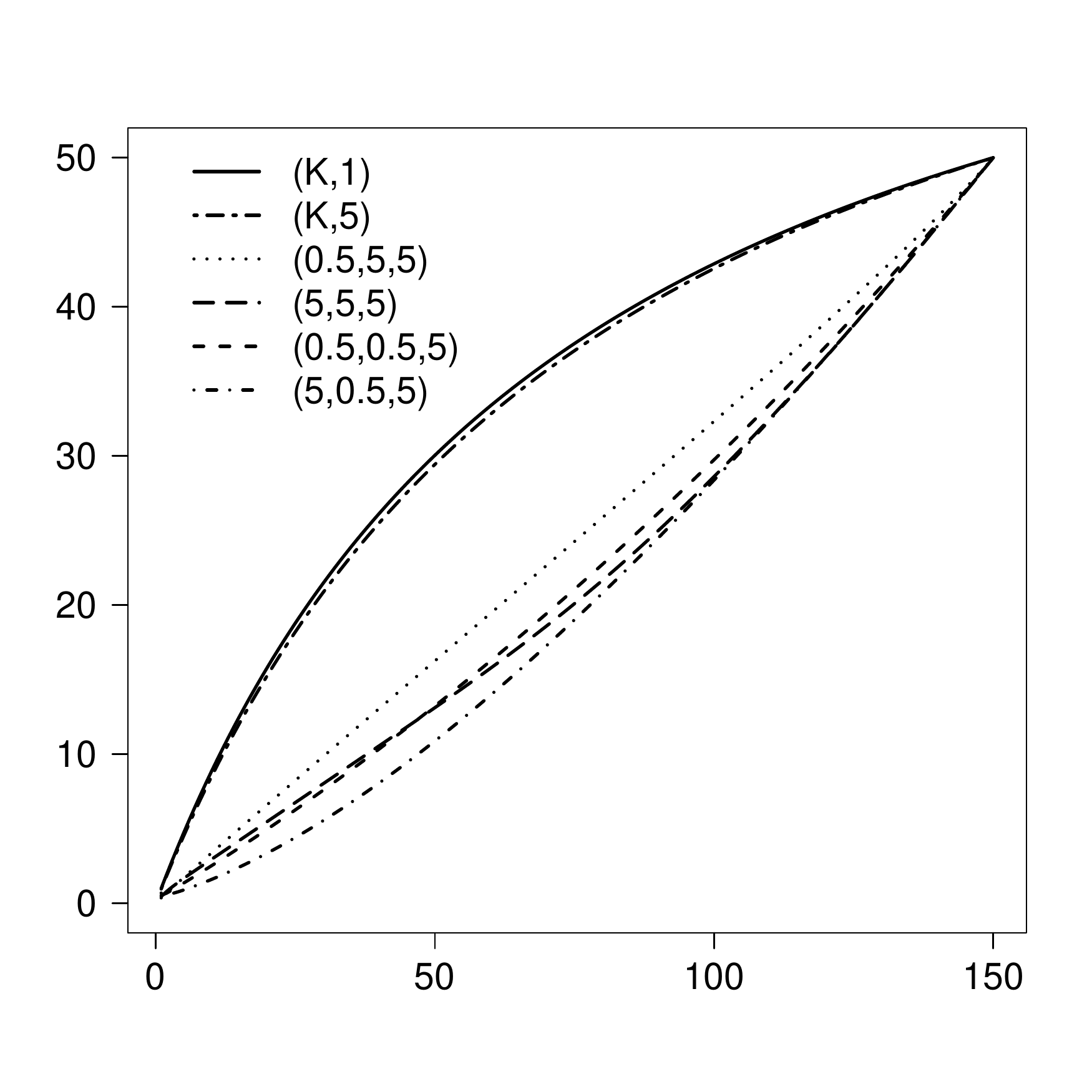}}%
  \put(60,-140){sample size $\sub{a}{1}$}\put(260,-140){sample size $\sub{a}{1}$}
  \put(60,-330){sample size $\sub{a}{1}$}\put(260,-330){sample size $\sub{a}{1}$}
  \end{picture}%

\clearpage
\pagebreak
\newpage

Figure 3: Expected value $\mathbb{E}[\sub{A}{1}^*(\sub{a}{2}) = \sub{a}{1} | \sub{A}{1}(0) =
150, \sub{A}{2}(0) = 50]$ as a function of the number $\sub{a}{2}$ of
ancestral lineages belonging to the subsample varying over $\pi$ and
$\theta$ as shown in the legends. By $K$ we denote the Kingman coalescent.

\begin{picture}(50,50)
  \put(0,-140){\includegraphics[height=2.5in,width=2.5in,scale=1,angle=0]{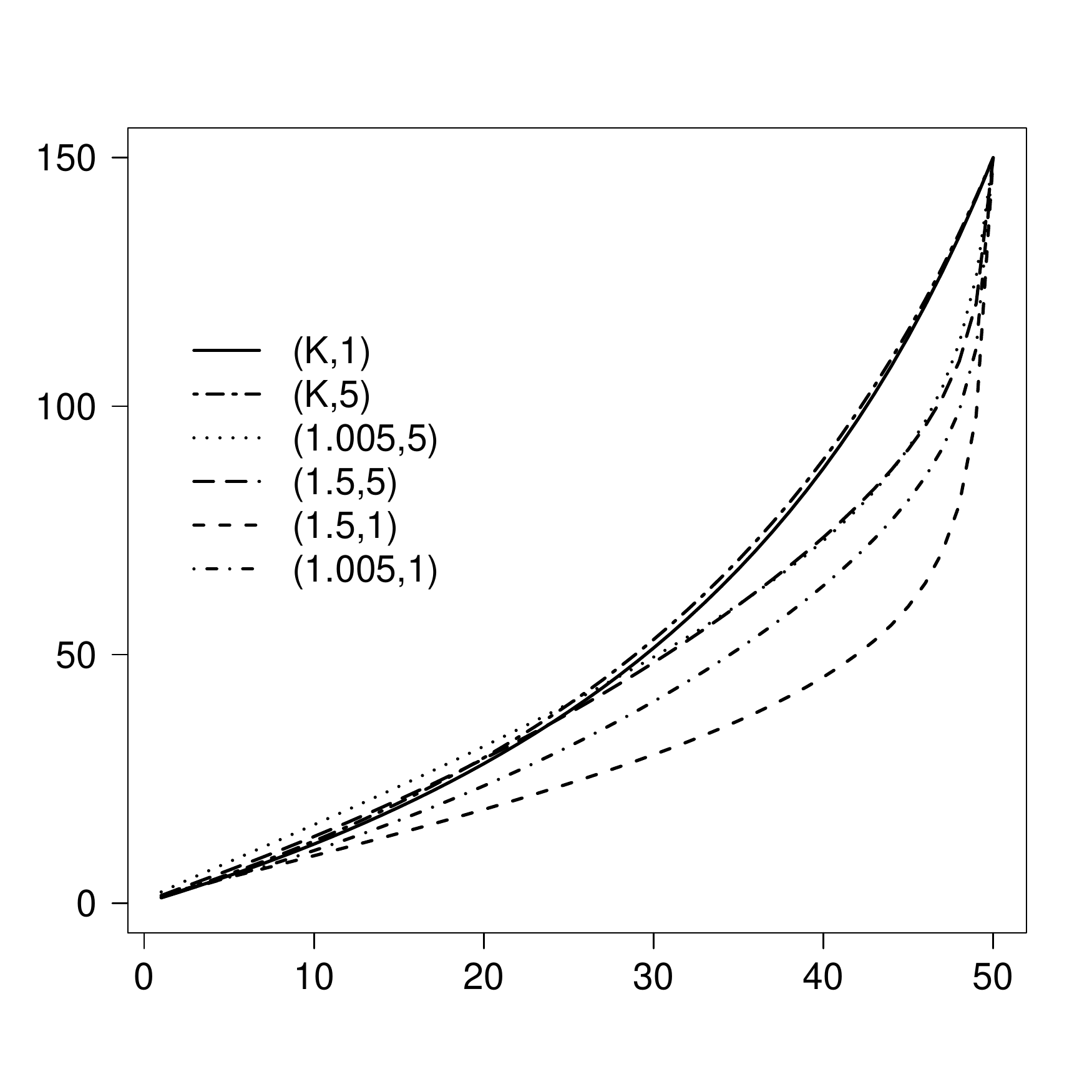}}%
   \put(0,-330){\includegraphics[height=2.5in,width=2.5in,scale=1,angle=0]{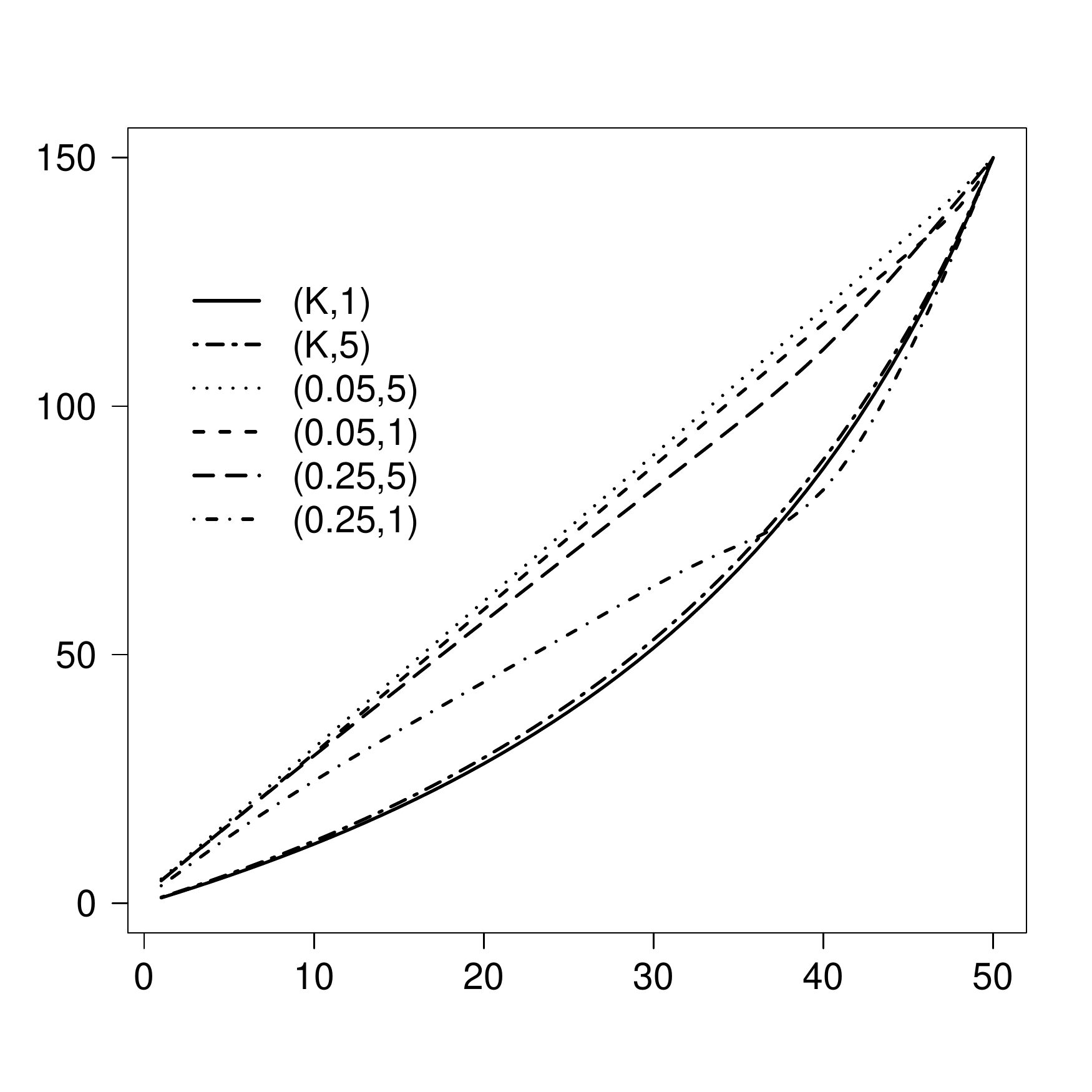}}%
   \put(200,-140){\includegraphics[height=2.5in,width=2.5in,scale=1,angle=0]{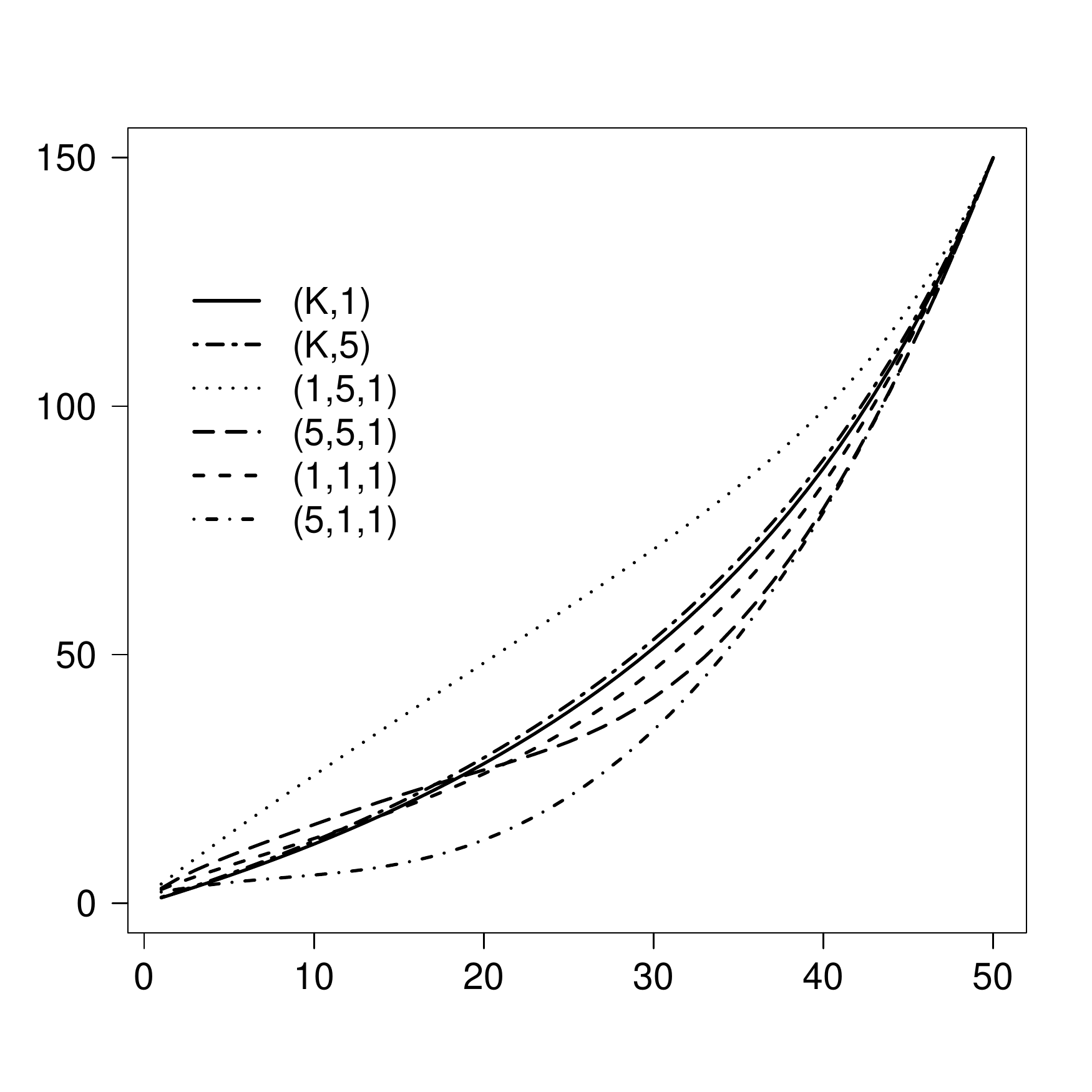}}%
   \put(200,-330){\includegraphics[height=2.5in,width=2.5in,scale=1,angle=0]{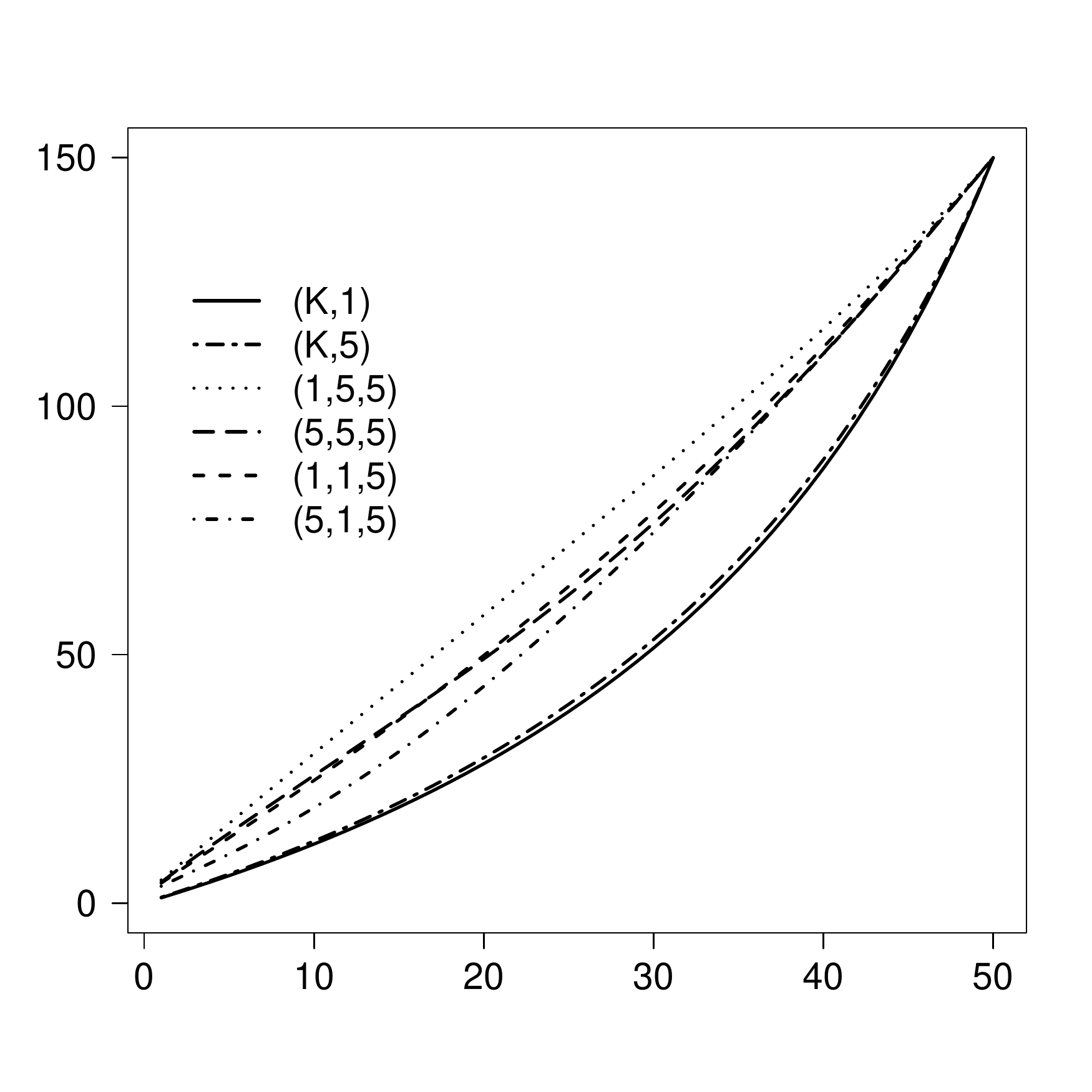}}%
   \put(50,25){beta coalescent}\put(250,25){beta $(\alpha,\beta)$ coalescent}%
   \put(50,-165){point mass coalescent}\put(250,-165){beta $(\alpha,\beta)$ coalescent}%
    \put(60,-140){sample size $\sub{a}{2}$}\put(260,-140){sample size $\sub{a}{2}$}
  \put(60,-330){sample size $\sub{a}{2}$}\put(260,-330){sample size $\sub{a}{2}$}%
  \put(30,-10){$(\alpha,\theta)$}\put(230,0){$(\alpha, \beta, \theta)$}%
   \put(30,-190){$(\psi,\theta)$}\put(230,-190){$(\alpha, \beta, \theta)$}%
  \end{picture}%

\clearpage
\pagebreak
\newpage

Figure 4: The probability $\tilde{u}(150,j)$ that the oldest allele of
a sample is among $j$ lineages as a function of $j$ from 1 to 50 and
varying over $(\pi,\theta)$ as shown in the legends (in the same form
as in Figure~3). By $K$ we denote the usual Kingman coalescent. 

\begin{picture}(50,50)
  \put(0,-140){\includegraphics[height=2.5in,width=2.5in,scale=1,angle=0]{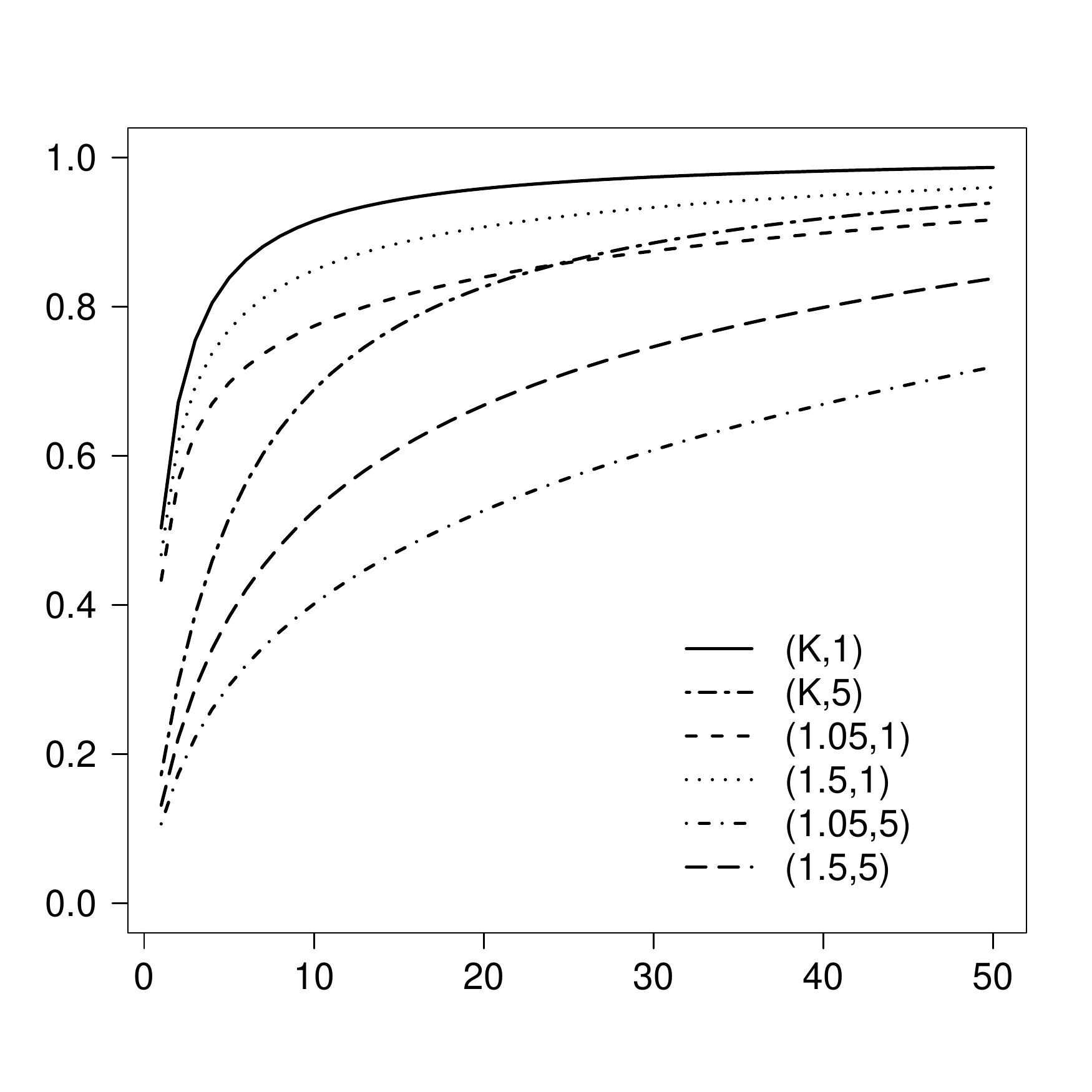}}%
   \put(0,-330){\includegraphics[height=2.5in,width=2.5in,scale=1,angle=0]{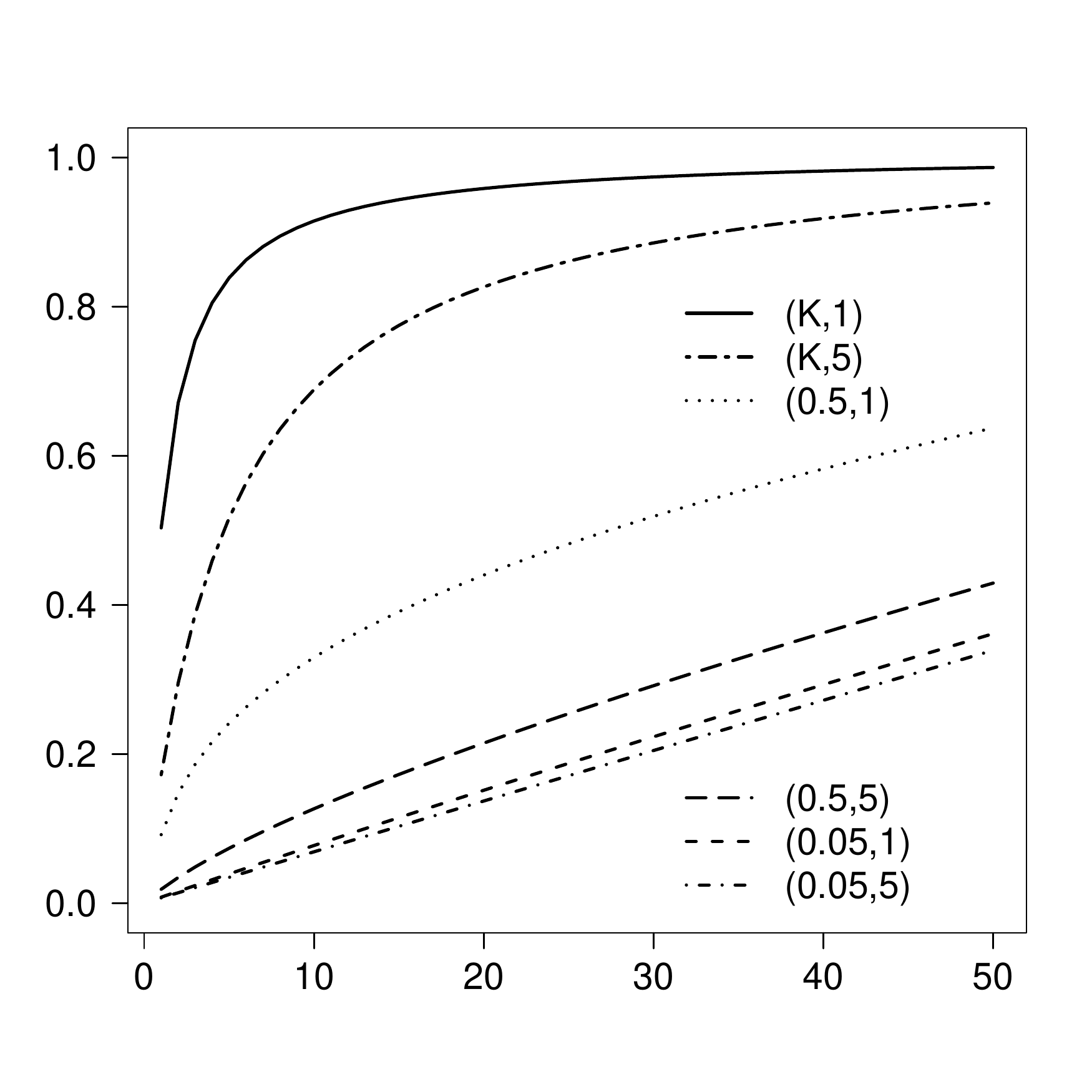}}%
   \put(200,-140){\includegraphics[height=2.5in,width=2.5in,scale=1,angle=0]{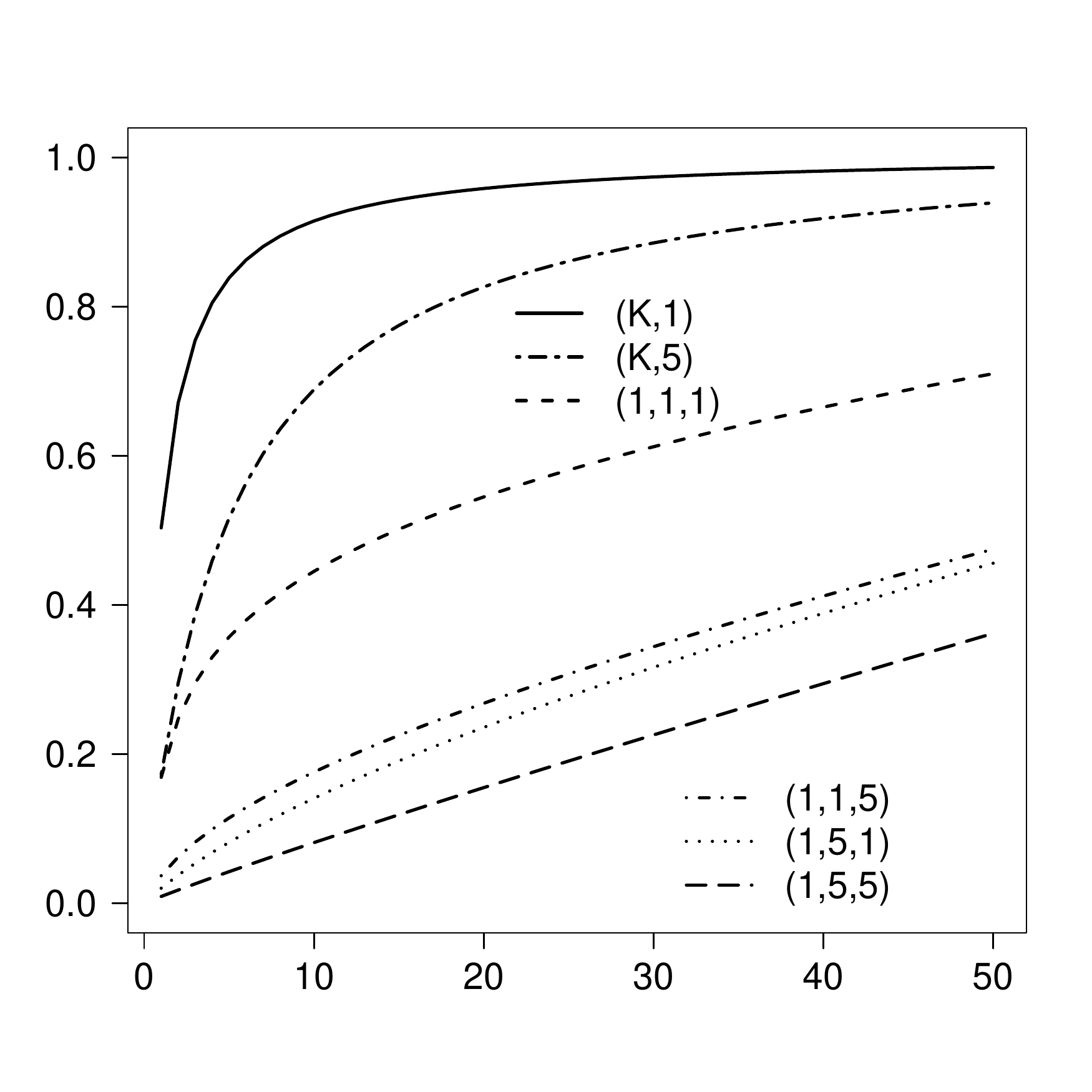}}%
   \put(200,-330){\includegraphics[height=2.5in,width=2.5in,scale=1,angle=0]{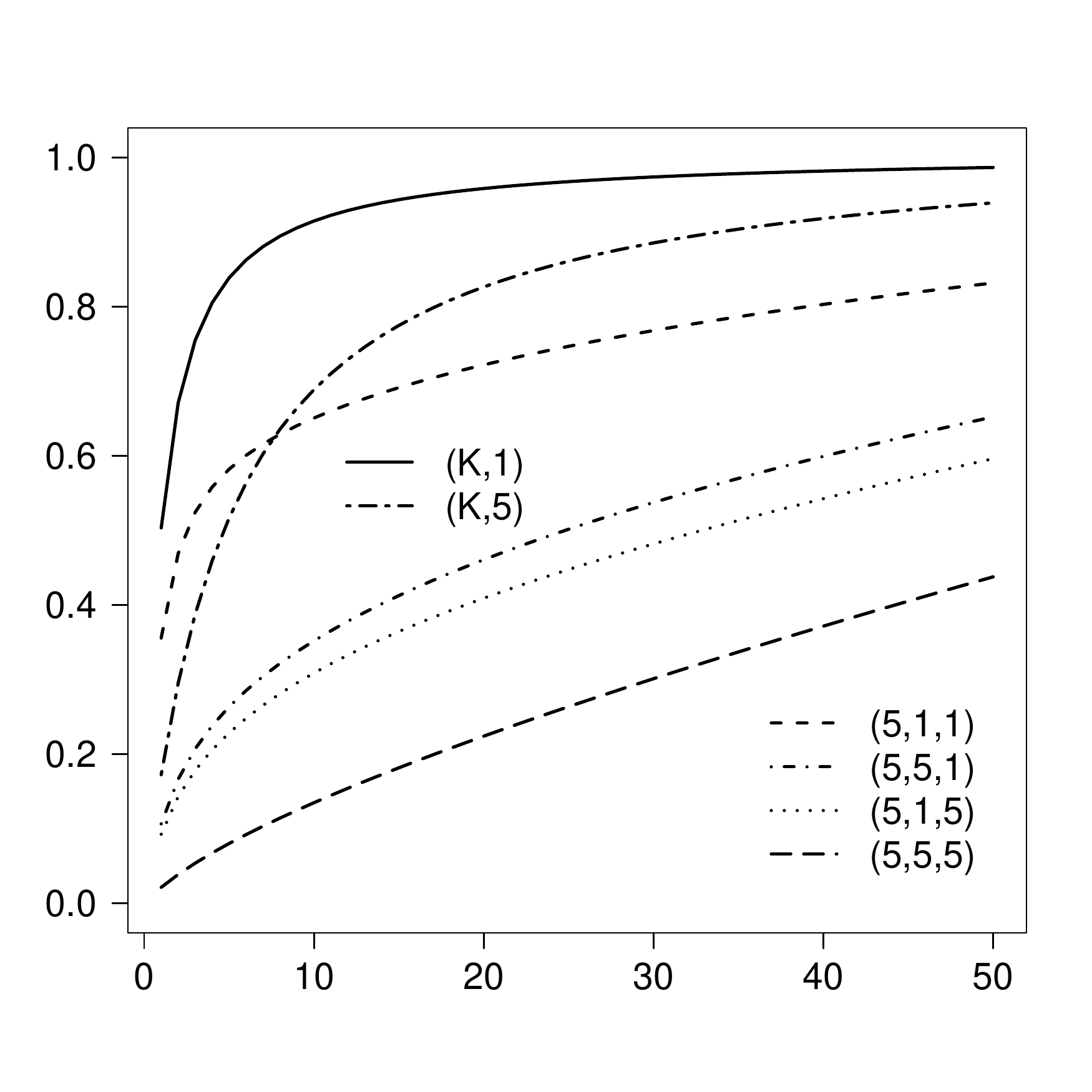}}%
    \put(50,25){beta coalescent}\put(250,25){beta $(\alpha,\beta)$ coalescent}%
   \put(50,-165){point mass coalescent}\put(250,-165){beta $(\alpha,\beta)$ coalescent}%
    \put(60,-140){subsample size $j$}\put(260,-140){subsample size $j$}
  \put(60,-330){subsample size $j$}\put(260,-330){subsample size $j$}%
  \end{picture}%

\end{document}